\documentclass[acmsmall,10pt,screen,nonacm]{acmart}
\setcopyright{none}
\acmDOI{}
\acmISBN{}
\acmConference{}{}{}
\acmBooktitle{}
\acmVolume{}
\acmNumber{}
\acmArticle{}
\acmYear{}
\copyrightyear{}
\usepackage{tabularx}
\usepackage{dcolumn}
\newcolumntype{d}[1]{D{.}{.}{#1}}
\usepackage{subcaption}
\usepackage{placeins}
\usepackage{enumitem}
\usepackage{xspace}
\usepackage{cleveref}
\usepackage{iftex}
\usepackage{pifont}
\AtBeginDocument{\renewcommand{\checkmark}{\ding{51}}}
\ifXeTeX
  \DeclareGraphicsExtensions{.pdf,.png,.jpg,.jpeg}
\else
  \DeclareGraphicsExtensions{.png,.jpg,.jpeg,.pdf}
\fi
\AtBeginDocument{}
\newcommand{\AI}{AI model alone\xspace}
\newcommand{\Hum}{Human alone\xspace}
\newcommand{\HAI}{Human+AI\xspace}

\newcommand{\todo}[2][]{}
\newcommand{\itodo}[1]{}
\newcommand{\donebot}[1]{}
\renewcommand{\bot}[1]{}
\newcommand{\michelle}[1]{}
\newcommand{\daniela}[1]{}
\newcommand{\robin}[1]{}
\newcommand{\agnes}[1]{}
\newcommand{\rw}[1]{}
\makeatletter
\let\hyxmp@parse@acmart\relax
\makeatother
\hypersetup{unicode=true,keeppdfinfo=true,colorlinks=true,linkcolor=blue,citecolor=blue,urlcolor=blue}
\fancypagestyle{preprintpagestyle}{%
  \fancyhf{}
  \fancyhead[L]{\small Confident, Not Wiser}
  \fancyhead[R]{\small Preprint}
  \fancyfoot[C]{\thepage}
  
}
\fancypagestyle{preprintfirstpagestyle}{%
  \fancyhf{}
  \fancyfoot[C]{\thepage}
  
}
\AtBeginDocument{\fancypagestyle{plain}{%
  \fancyhf{}
  \fancyhead[L]{\small Confident, Not Wiser}
  \fancyhead[R]{\small Preprint}
  \fancyfoot[C]{\thepage}
  
}}

\title[Confident, Not Wiser]{Confident, Not Wiser: The Dunning-Kruger Effect in Human-AI Interaction}
\author{Daniela Fernandes}
\orcid{0009-0006-1332-7485}
\affiliation{\institution{Aalto University}\city{Espoo}\country{Finland}}
\email{daniela.dasilvafernandes@aalto.fi}
\author{Michelle Rausch}
\orcid{0009-0002-7097-9675}
\affiliation{\institution{Aalto University}\city{Espoo}\country{Finland}}
\author{Agnes Mercedes Kloft}
\orcid{0009-0008-8024-2398}
\affiliation{\institution{Aalto University}\city{Espoo}\country{Finland}}
\email{agnes.kloft@aalto.fi}
\author{Daniel Buschek}
\orcid{0000-0002-0013-715X}
\affiliation{\institution{University of Bayreuth}\city{Bayreuth}\country{Germany}}
\email{daniel.buschek@uni-bayreuth.de}
\author{Robin Welsch}
\orcid{0000-0002-7255-7890}
\affiliation{\institution{Aalto University}\city{Espoo}\country{Finland}}
\email{robin.welsch@aalto.fi}

\hypersetup{pdfauthor={Daniela Fernandes, Michelle Rausch, Agnes Mercedes Kloft, Daniel Buschek, Robin Welsch},pdfsubject={Research preprint}}
\date{15 September 2026}
\begin{document}
\begin{abstract}
AI assistance can improve performance without improving self-assessment. 
We report a study (N=366) comparing Human alone and Human+AI performance on reasoning tasks, for which the AI model is benchmarked on the same items. Participants estimated global and block performance and rated confidence in their answers. Human+AI achieved higher scores, but self-estimates tracked performance weakly. Average overestimation was similar across groups, covering individual errors. Across tasks, confidence distinguished correct from incorrect answers less accurately in the Human+AI group, while within-task differences remained uncertain. The Dunning--Kruger pattern was found in both groups, with a larger observed contrast in Human+AI. Controls for score noise reduced but did not eliminate the pattern, with the controlled group difference remaining inconclusive. An extended computational account describes global and block estimates. Our findings distinguish performance augmentation from metacognitive augmentation and motivate interfaces that support verification, communicate task-specific AI model performance, and help users evaluate the quality of their joint work rather than produce answers.
\end{abstract}
\keywords{metacognition, Dunning-Kruger effect, calibration, confidence, large language models, human-AI interaction}
\maketitle
\pagestyle{preprintpagestyle}
\thispagestyle{preprintfirstpagestyle}

\begin{figure}[htbp]
\centering

  \includegraphics[width=\textwidth]{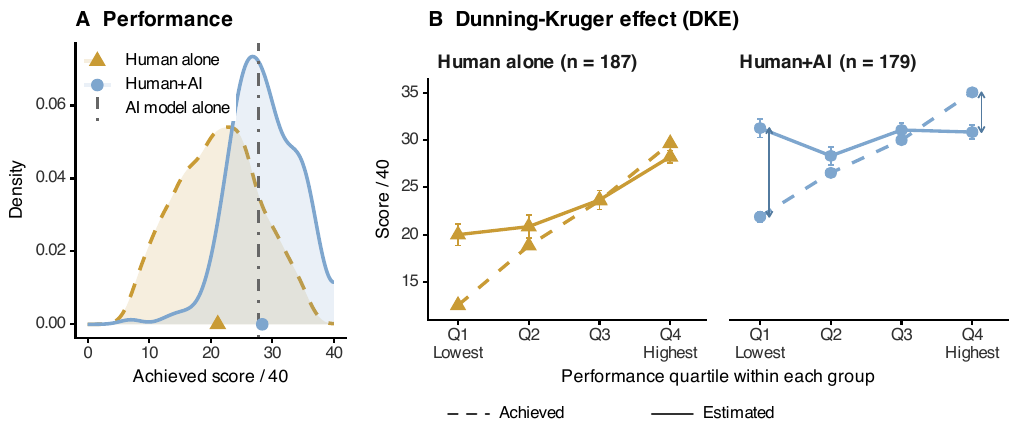}
  \caption{Higher performance does not guarantee better self-assessment. (A) Human+AI scored higher than Human alone, while the overall contrast with the AI model alone benchmark remained uncertain. (B) Across within-group performance quartiles, Human+AI estimates stayed near 30/40 while achieved scores varied substantially. Arrows highlight overestimation among the lowest scorers and underestimation among the highest. Yellow denotes Human alone and blue Human+AI; the grey line marks the AI model benchmark. Dashed lines show achieved scores, solid lines estimates, and error bars $\pm1$ SEM. These are descriptive associations; score-noise controls are reported in the Results.}
  \Description{Left: Yellow and blue distributions show higher achieved scores for Human+AI. Right: panels labeled Human alone (n = 187) and Human+AI (n = 179) compare achieved and estimated scores across four performance quartiles under the heading Dunning-Kruger effect (DKE). Human+AI estimates remain near thirty while achieved scores rise from roughly twenty-two to thirty-five. Arrows mark overestimation at the lowest and underestimation at the highest Human+AI quartile.}
  \label{fig:teaser}
\end{figure}
\section{Introduction}

Users are increasingly solving problems with AI models, particularly large language models (LLMs), and as they do so, they must consider what they create on their own, what the AI model produces, and what the two can accomplish together. The AI model's performance varies and is not directly visible to users. It varies from near-certain to chance level within the same kind of problem \cite{mccoy2024embers}, and the AI model gives few signals of its own uncertainty \cite{zhou2024reliable}. The judgments users must make are therefore metacognitive (i.e., judgments about one's own thinking, here extended to thinking with AI \cite{fleming2024metacognition, tankelevitch2023metacognitive}).

Metacognition captures two distinct measures that can move independently. \textit{Metacognitive accuracy} corresponds to the difference between a person's estimate of their own performance and the performance they actually achieved. \textit{Metacognitive sensitivity} corresponds to how well the confidence attached to individual answers distinguishes between correct and incorrect answers \cite{fleming_how_2014, fiedler2019metacognition}. 
Both contribute to whether the Human+AI pair gains from working together. When human and AI model errors differ, above-chance discrimination between correct and incorrect answers can allow for an optimal combination where it outperforms both members \cite{li2026metacognitive}. When this discrimination fails, overreliance \cite{lee2004trust} can undermine the human contribution so that Human+AI performs worse than Human alone \cite{bastani2024generative, vaccaro2024combinations}. A user working with an AI model therefore has to evaluate, answer by answer, whether to accept what the AI model produced. AI assistance thus changes both the performance being judged and the evidence available for judging it, and how access to an AI model relates to metacognition is unclear.

Self-judgments of this nature are known to be inflated, both regarding one's own performance and what AI adds to it. People estimate themselves to be better than average \cite{brown1986evaluations, zell2020better}, believe that AI improves their performance \cite{kloft2023ai, cave2019hopes}, and overestimate AI gains even from a sham AI system \cite{villa2023placebo, kosch2023placebo}. The best-known version of this failure is the Dunning-Kruger Effect (DKE), in which low performers overestimate their performance and high performers underestimate it \cite{kruger1999unskilled, dunning2011dunning}. Note that regression to the mean and measurement error reproduce the same pattern without any actual deficit in self-monitoring \cite{krueger2002unskilled, gignac2020dunning, jansen2021rational}. 

How people monitor an AI model is even less understood. Reliance does not reliably track AI model accuracy \cite{lu2021human, klingbeil2024trust}, and users misjudge output quality \cite{kelly2023capturing, fernandes_ai_2024}. Closest to our work, \citet{fernandes_ai_2024} found that assistance from an AI model raised reasoning performance but left participants overestimating their performance. They concluded that the DKE disappeared with AI assistance because their computational criterion was not met, although lower-performing AI users still overestimated more than higher performers. Whether this computational conclusion generalizes across tasks remains unresolved. We extend this work across four reasoning task types using global estimates, block estimates, and item confidence to examine performance-dependent estimation errors and distinguish pooled from within-task discrimination.

We report an exploratory observational study in which separately recruited \textit{Human alone} and \textit{Human+AI} groups completed reasoning tasks spanning matrices, mental rotation, syllogisms, and letter strings. We benchmarked the AI model alone on the same items, placing beliefs about the self, the AI model, and the Human+AI pair on a common scale \cite{klein2024performance}. The tasks were selected to vary AI model performance. Participants rated confidence in individual answers and estimated performance globally and by task block. The study was not preregistered. Three research questions organize the analyses:

\begin{itemize}
\item \textbf{RQ1}: Does Human+AI outperform the AI model alone, or only Human alone, and how accurately do participants estimate their own solo performance, AI model performance, and Human+AI performance?
\item \textbf{RQ2}: How do metacognitive accuracy (calibration) and metacognitive sensitivity (discrimination) differ between Human alone and Human+AI?
\item \textbf{RQ3}: Does the Dunning-Kruger effect survive a regression-to-the-mean control and a correction for measurement error, and how is its performance-dependent component distributed between item-confidence error and the discrepancy between summed confidence and the global estimate?
\end{itemize}

We find that, although the Human+AI group outperformed the Human alone group, their self-estimates did not accurately track performance. The DKE pattern remained in both groups after controls for score noise. Together, these findings distinguish performance augmentation from metacognitive augmentation and thus motivate support for self-assessment during interaction with AI models.

We make three contributions. First, we examine metacognitive accuracy and sensitivity together through global estimates, task-level estimates, and confidence in individual answers, distinguishing pooled from within-task discrimination. Second, we apply DKE controls to human--AI collaboration to assess which patterns persist after accounting for specific measurement artifacts. Third, we derive design implications for supporting assessments of Human+AI performance alongside task-specific AI model performance.

\section{Related Work}

We draw on two lines of research: The first is metacognition, which separates assessing one's own total performance from assessing one's own individual answers. We carry this separation through the paper as two outcomes which prior work on human-AI interaction suggests may be affected differently. The second is the DKE, with the statistical robustness checks and computational accounts of the phenomenon.

\subsection{Metacognition in Human-AI Interaction}
Evaluating one's own performance involves two distinct quantities, which we treat as separate outcomes. Signed calibration error corresponds to the difference between perceived and actual performance~\cite{fiedler2019metacognition, fleming2024metacognition}, e.g., how many items on a test one judges to have solved correctly versus the test score. It fails through bias, a consistent error in one direction~\cite{fleming2024metacognition}, or through noise, random fluctuation from one occasion to the next. Metacognitive sensitivity, or discrimination, is the ability to distinguish one's correct from incorrect responses, measured from confidence judgments given after each decision~\cite{fleming_how_2014}. A participant who estimates their total almost exactly may still be unable to say which answers were correct. Both vary considerably between individuals~\cite{kelemen2000individual} and depend on which cues a judgment draws on~\cite{koriat1997monitoring, ackerman2017meta}. In signal detection terms, a confidence rating is a second-order decision about whether one's first-order response was correct, so sensitivity can be quantified as the detectability of one's own errors~\cite{maniscalco2012signal, fleming_how_2014}.
\textbf{We measure calibration from participants' estimates of their score and sensitivity from their confidence on every item, and treat them as separate outcomes that may differ independently between Human alone and Human+AI.}

Metacognitive sensitivity is commonly quantified as the area under a participant's type-2 ROC curve, where~.5 means confidence carries no information (chance level) and 1 means perfect discrimination~\cite{fleming_how_2014}; for an overview of indices, see~\citet{rahnev2025comprehensive}. Metacognitive sensitivity also limits the performance the Human+AI pair can achieve. Under suitable conditions on the dependence of human and AI model errors, informative confidence from either agent can allow an optimal combination of their predictions to outperform both members, even when AI model performance is higher \cite{li2026metacognitive}.
\textbf{This study compares sensitivity in Human alone and Human+AI to examine its role in the potential benefits of AI assistance (RQ2).}

AI assistance can improve performance on reasoning and decision tasks~\cite{steyvers_bayesian_2022, 10.1145/3411764.3445717, tankelevitch2025understanding}, the goal of the field since~\citet{engelbart1962augmenting} framed computing as amplifying intellect. \citet{vaccaro2024combinations} distinguish \textit{synergy} (Human+AI outperforms both Human alone and the AI model alone) from \textit{augmentation} (Human+AI outperforms Human alone) and find synergy rare when the AI model is the stronger of the two. Neither outcome indicates whether the person is still aware of how well they performed. A score obtained using an AI model may be higher even when some reasoning is supplied by the AI model, so the cues that a self-judgment typically relies on may be missing or misleading. Moreover, self-assessment draws on internal cues and external feedback~\cite{koriat1997monitoring, ackerman2017meta}. When assistance is immediate and effortless, the source of an answer is easily forgotten, and solutions achieved with AI are credited to oneself~\cite{johnson_source_1993, Zindulka2026AIMemoryGap, draxlerghost24}. Fluent explanations inflate the sense of understanding~\cite{rozenblit2002misunderstood}, outside assistance raises estimates of what one could achieve \emph{without} it~\cite{fisher2021harder}, and system explanations do not reliably correct this. Users neglect them~\cite{wang2021explanations} or shift trust to explanations that carry no information~\cite{eiband2019impact, bertrand2022cognitive}.
\textbf{We measure performance and self-assessment in the same participants within each group and compare both outcomes between Human alone and Human+AI.}

These mechanisms surface as overreliance. Operators of imperfect automation stop monitoring its output~\cite{wickens2015complacency}, early impressions anchor later reliance~\cite{10.1145/3397481.3450639}, and reliance is hardest to withhold when the AI model is confidently mistaken~\cite{si2024large}. Interfaces that force deliberate engagement or state the AI model's uncertainty reduce overreliance~\cite{kim2024m, forcing2021, vasconcelos2023explanations}, so part of the deficit is designed into the interaction. Beliefs about assisted performance are inflated while outcomes go unchecked~\cite{colombatto2025metacognition, kloft2023ai, von2025knowing}, shift even when assistance is never delivered~\cite{kosch2023placebo, villa2023placebo}, and credit the person with gains that belong to the pair~\cite{stadler_cognitive_2024, bastani2024generative, rafner2022deskilling, Kobiella2024}. How users judge AI model performance remains poorly understood. Reliance does not reliably track AI model accuracy~\cite{lu2021human, okamura2020empirical, klingbeil2024trust}; an inflated view of one's own competence obstructs appropriate reliance~\cite{he2023knowing}; and letting users compare their correctness with the AI model's improves the pair's decisions~\cite{ma2023who}. The monitoring constraint is metacognitive rather than informational~\cite{tankelevitch2023metacognitive}, and interfaces are rarely designed to reduce it~\cite{ramesh2026metacognitive}.
\textbf{We compare beliefs about the self with observed human scores and beliefs about the AI model with the AI model alone benchmark on the same items, so that misjudging the AI model is measured directly rather than inferred from reliance.}

Beliefs about Human+AI are the least studied of the three. People credit algorithms with more accuracy than human experts~\cite{shekar2024people, logg2019algorithm}, abandon them once they have seen them fail ~\cite{dietvorst2015algorithm}, and expect AI to improve their own performance~\cite{kloft2023ai, cave2019hopes}. Whether people expect Human+AI to exceed its better member, the synergy that~\citet{vaccaro2024combinations} find rare, has to our knowledge not been set against what the pair actually achieves.
\textbf{We elicit beliefs about Human+AI alongside beliefs about the self and the AI model, which lets us compare expected synergy with observed group performance and the AI model alone benchmark (RQ1).}

\subsection{The Dunning-Kruger Effect}
The DKE links the accuracy of self-assessment to individual skill: low performers overestimate and high performers underestimate their performance~\cite{kruger1999unskilled}. Large-scale studies have reproduced it~\cite{ehrlinger2008unskilled, jansen2021rational}, and the effect has been reported well beyond its original tasks~\cite{dunning2011dunning, yang2024competence}. Whether the effect requires a metacognitive explanation is contested. Regression to the mean combined with the better-than-average effect reproduces the pattern, as participants are ranked on the same score that enters the estimate--performance difference and self-evaluations sit above the midpoint regardless of skill~\cite{krueger2002unskilled, zell2020better}. Moreover, so does measurement noise alone~\cite{nuhfer2016random}. An observed score is itself an imperfect measure of ability, and this noise flattens the relation between estimate and performance. \citet{feld2017estimating} correct for this with an independent measure of the same ability, and~\citet{gignac2020dunning} offer two diagnostics: curvature and unequal residual spread can challenge a linear, constant-noise account, although neither diagnostic excludes all statistical explanations. Task difficulty offers a further test, since the effect reverses on hard tasks~\cite{burson2006skilled}, which a satisfactory account needs to accommodate. More generally, people overestimate on hard tasks and underestimate on easy ones~\cite{moore2008trouble}, and overestimation should be read against the difficulty of the task it was measured on. These controls are rarely applied together.
\textbf{We apply the grouping control, the measurement-error correction, and both diagnostics to Human alone and Human+AI alike and read block-level overestimation against block difficulty to assess which aspects of the DKE persist after these checks (RQ3)}.

Two computational models provide different accounts of the effect. The rational-observer account~\cite{jansen2021rational} treats a self-estimate as an inference about one's ability from noisy observations of one's own attempts. The posterior is pulled toward the prior mean, with weak and strong performers in opposite directions, and the DKE pattern appears without any self-monitoring deficit. In this computational model, the probability $\varepsilon$ of being wrong about whether an answer was correct is capped at~.5 and parameterizes sensitivity. The bias-and-noise computational model of~\citet{fernandes_ai_2024} instead places the effect in the mapping from achieved to reported performance. A bias term shifts each report in one direction, while a noise term scales how strongly reports follow skill, with compression allowing a declining estimation-error pattern across skill. The accounts locate the effect in inference versus in reporting, and therefore differ in what AI assistance could alter. We extend the second computational model to block-level estimates, which jointly represent bias, compression, block difficulty, and person-specific report variation.
\textbf{We decompose overestimation into confidence-minus-score and global-minus-summed-confidence components to describe their associations with performance in each group, without relying on either fit (RQ3).}

The status of the DKE in human--AI interaction remains unresolved.
\citet{fernandes_ai_2024} reported greater overestimation among lower-performing AI users, but their computational DKE criterion was not met. This could reflect reduced performance differences, a common upward bias combined with score noise, or constraints imposed by task difficulty and bounded global estimates.

We examine these alternatives across four reasoning tasks with varying AI model performance, using score-noise controls and an extension of their computational account. A same-item AI model benchmark anchors beliefs about the self, the AI model, and Human+AI to measured performance.
\textbf{Global and block estimates let us test whether performance-dependent estimation errors persist across tasks and reporting levels under these controls.}

\section{Method}
\label{sec:design}

Participants completed a purpose-built 40-item reasoning battery either on their own (\Hum group) or alongside a conversational AI model (\HAI group; GPT-5.6 Luna), and estimated their performance before, during, and after the battery. The public release location for this paper's research software, data, and analysis scripts is pending author confirmation. This study did not require ethics review under Aalto University's criteria for mandatory review, which follow the Finnish National Board on Research Integrity (TENK) guidelines. \label{sec:method:availability}

\subsection{Participants}
\label{sec:method:participants}

We recruited participants through separate calls on Prolific. Each participant was assigned to the group whose call they responded to first. The application then made them automatically ineligible for participation in the other group. We restricted participation to legally competent adults aged 18 or older who reported English as their first language and the United Kingdom as their country of residence. Participants received \pounds 9 per hour and performance bonuses. Performance bonuses were framed as incentives. Participants were ranked according to their scores on a leaderboard (one for each group) and were informed beforehand that the top ten would each earn a bonus equivalent to half the hourly wage, with the highest scorer receiving an additional \pounds 100.

An a priori power analysis for a two-sample $t$-test (smallest effect of interest $d = 0.30$, $\alpha = .05$, two-sided, power $= .80$) required $176$ participants per group. We excluded sessions on three criteria: incomplete sessions, failure of at least two of the three attention checks, and proctoring events such as consulting an external AI model or extended periods offline (\autoref{sec:method:apparatus}). The proctoring service flagged sessions automatically, but a flag did not lead to exclusion on its own. We reviewed each flagged session and decided case by case. We applied these criteria continuously while data collection was running and replaced excluded sessions by further recruitment until both groups passed the target, so we do not report a single recruited total from which the analyzed sample was subtracted.\footnote{Our record of removals lists 65 participants excluded after manual review of proctoring flags, 23 for proctoring alone and 42 for incomplete sessions. None of the 65 appear in the final sample. Prolific monetary compensations were independent of analysis inclusion. Note that this record covers only the removals due to proctoring flags, so these counts do not sum to the full recruitment sample.} Our final sample consisted of $N=366$: 187 in \Hum\ and 179 in \HAI\ (demographics in \autoref{tab:demographics}).

\begin{table}[h!]
\centering
\caption{Participant demographics by group}
\label{tab:demographics}
\small
\begin{tabular}{lrrr}
\toprule
 & Human alone ($N$=187) & Human+AI ($N$=179) & Total ($N$=366) \\
\midrule
Age, $M$ (SD)         & 36.8 (10.9) & 34.3 (10.8) & 35.6 (10.9) \\
Age range              & 18--67 & 18--74 & 18--74 \\
\midrule
\multicolumn{4}{l}{\textit{Gender, $n$ (\%)}} \\
\quad Man                              & 99 (52.9)  & 97 (54.2)  & 196 (53.6) \\
\quad Woman                            & 87 (46.5)  & 80 (44.7)  & 167 (45.6) \\
\quad Non-binary / multiple categories & 1 (0.5)    & 1 (0.6)    & 2 (0.5) \\
\quad Prefer not to disclose           & 0 (0.0)    & 1 (0.6)    & 1 (0.3) \\
\midrule
\multicolumn{4}{l}{\textit{Education, $n$ (\%)}} \\
\quad Basic education / middle school  & 24 (12.8)  & 20 (11.2)  & 44 (12.0) \\
\quad Vocational college degree        & 27 (14.4)  & 38 (21.2)  & 65 (17.8) \\
\quad Bachelor's level                 & 87 (46.5)  & 74 (41.3)  & 161 (44.0) \\
\quad Master's level                   & 40 (21.4)  & 42 (23.5)  & 82 (22.4) \\
\quad Doctoral degree                  & 9 (4.8)    & 5 (2.8)    & 14 (3.8) \\
\midrule
\multicolumn{4}{l}{\textit{AI use frequency, $n$ (\%)}} \\
\quad Never                            & 5 (2.7)    & 2 (1.1)    & 7 (1.9) \\
\quad A couple of times per year       & 8 (4.3)    & 7 (3.9)    & 15 (4.1) \\
\quad Once a month                     & 30 (16.0)  & 20 (11.2)  & 50 (13.7) \\
\quad Every week                       & 74 (39.6)  & 65 (36.3)  & 139 (38.0) \\
\quad Every day                        & 70 (37.4)  & 85 (47.5)  & 155 (42.3) \\
\bottomrule
\end{tabular}
\end{table}

\subsection{Study design and interface}
\label{sec:method:apparatus}

\begin{figure}[h]
\centering
 \includegraphics[width=\linewidth]{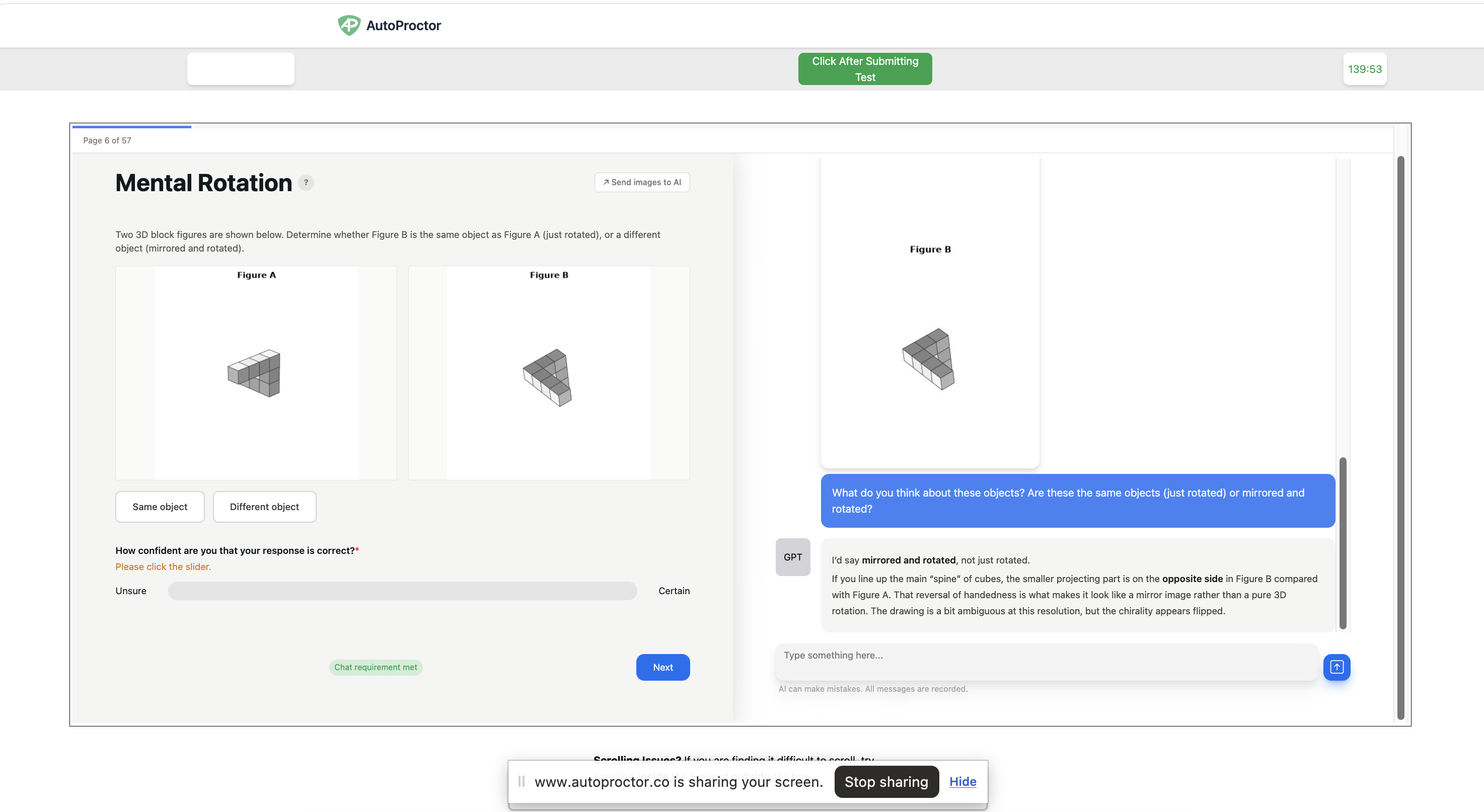}
\caption{
The study application used a split interface, with the reasoning task and the response and confidence controls on the left and the AI model on the right. In the Human alone group the same pages were shown without the chat panel.
}
\label{fig:interface}
\Description{Study interface showing the reasoning task, answer selection, and confidence controls beside an AI chat panel.}
\end{figure}

The two groups differed in access to a conversational AI model. Participants in
the \HAI\ group saw GPT-5.6 Luna in an AI model chat panel alongside every item and had to send at least one message to the AI model before submitting an answer (see \autoref{fig:interface}). Participants in the \Hum\ group worked through the same 40 items without the AI model chat panel. We additionally evaluated GPT-5.6 Luna on the same items without a participant, which gives the \AI\ reference against which we interpret joint performance (\autoref{sec:method:benchmark}).
Each task page showed the task item, response controls, and confidence slider on the left, and in \HAI\ a streaming chat panel on the right (\autoref{fig:interface}). We hid this panel on every questionnaire and self-assessment page, so participants could not ask the AI model how well they had done.

The battery and questionnaires ran inside AutoProctor, which monitors screen content, tab switches, and navigation away from the study. Interface implementation and logging details are in Appendix~\ref{app:procedure_detail}.

\subsection{Tasks and measures}
\label{sec:method:battery}
\begin{figure}[h!]
\centering
\begin{minipage}[t]{0.42\linewidth}
  \vspace{0pt}
  \centering
  \includegraphics[width=\linewidth]{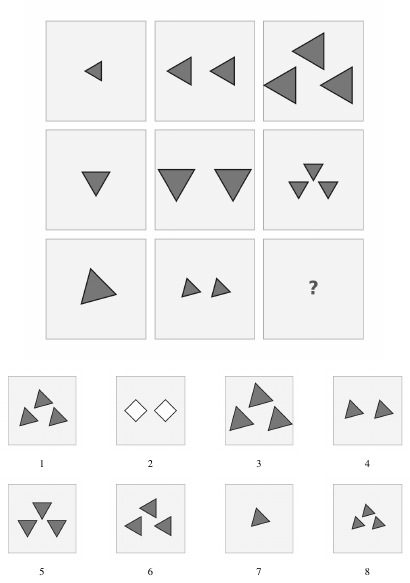}
  \par\smallskip
  {\footnotesize (a) Matrix reasoning: Complete the $3\times3$ pattern; answer options 1--8 below.}
\end{minipage}\hfill
\begin{minipage}[t]{0.53\linewidth}
  \vspace{0pt}
  \centering
  \includegraphics[width=\linewidth]{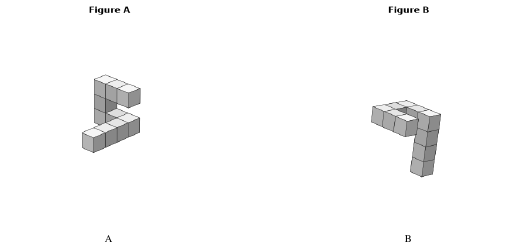}
  \par\smallskip
  {\footnotesize (b) Mental rotation: Is figure~B the same object as figure~A and just rotated, or its
  mirror image?}
  \par\bigskip
  \begin{flushleft}\footnotesize
  (c) Letter-string analogy (free-text answer):\par\medskip
  \scriptsize\linespread{1.3}\selectfont
  \texttt{Alphabet (in order): > * + < ! @ \$ ) \& = : - ( \% \textasciitilde}\par
  \texttt{Study the following patterns:}\par
  \texttt{[@ ) = -] $\rightarrow$ [@ ) = - \%]}\par
  \texttt{[\$ \& : (] $\rightarrow$ [\$ \& : ( \textasciitilde]}\par
  \texttt{Using the same alphabet, complete this:}\par
  \texttt{[> + ! \$] $\rightarrow$ ?}
  \end{flushleft}
\end{minipage}
\par\medskip
\noindent\rule{\linewidth}{0.4pt}\par\smallskip
\begin{minipage}{\linewidth}
\footnotesize (d) Syllogistic reasoning --- the conclusion is judged \emph{Valid} /
\emph{Invalid} / \emph{Cannot be determined}:\par\smallskip
\scriptsize\linespread{1.15}\selectfont
``At Dunmore Bakery, the manager keeps a daily log of all baked goods and their status. The log
shows the following:\par\medskip
The daily log notes that details for items logged by the opening team were last reviewed following
the most recent stock check. The bakery's records note that the number of items recorded under the
seasonal range code was verified at the end of last week. The record-keeping system is shared
across all staff on duty.\par\medskip
The manager reviews the bakery's records at the start of each week. Information relating to items
logged by the opening team is filed in a dedicated section of the bakery's daily log. Whenever an
entry in the bakery's records is classified as items logged by the opening team, it is also
classified as items assigned to batch code 3.\par\medskip
Among the bakery's records, there are entries classified as items assigned to batch code 3 that do
not carry the items recorded under the seasonal range code classification. The bakery's records
confirm that the list of items assigned to batch code 3 is updated at the start of each week. The
daily records are filed at the end of each shift by the duty manager.\par\medskip
\emph{Conclusion:} All items logged by the opening team are items recorded under the seasonal
range code.''
\end{minipage}
\par\smallskip\noindent\rule{\linewidth}{0.4pt}
\caption{One example item per task type, with the designed difficulty level and the correct answer given in parentheses:  (a)~Matrix reasoning (medium; option~1), (b)~Mental-rotation item (medium; \emph{same object}), (c)~Letter-string analogy (hard; \texttt{> + ! \$ \&}), and (d)~Syllogistic reasoning (hard, three mental models; \emph{cannot be determined}). In~(d) premises are embedded in a record-keeping narrative, so the quantifiers must be inferred from the prose.}
\label{fig:example-items}
\Description{Example matrix reasoning, mental rotation, letter-string analogy, and syllogistic reasoning items, illustrating the four task formats.}
\end{figure}

Participants completed four blocks containing ten items. One block each for matrix reasoning, mental rotation, syllogistic reasoning, and letter-string analogies (\autoref{fig:example-items}). Block and item order were randomized, with the questions of a syllogism scenario kept together. We generated items rather than sourcing published instruments that might occur in the AI model's training data. Construction rules and designed difficulty levels are documented in Appendices~\ref{app:generators} and~\ref{app:item_difficulty}.

\paragraph{Matrix reasoning}
Each item is a $3\times3$ grid of geometric figures in which the bottom-right image is replaced by a question mark. Participants completed the pattern by selecting one of eight answer images (Appendix~\ref{app:gen:mat}). The design follows Raven-like rule structures~\cite{carpenter_what_1990,Matzen2010}; related Sandia item sets have undergone psychometric evaluation~\cite{harris_measuring_2020}. Matrix analogies are also used in AI model benchmarks~\cite{pmlr-v80-barrett18a,Zhang_RAVEN}.

\paragraph{Mental rotation}
Following \citet{Shepard1971}, each item shows two three-dimensional block figures, each a chain of ten cubes. Participants judged whether figure~B was figure~A rotated (``same object'') or its mirror image (``different object'') (Appendix~\ref{app:gen:rot}). Related mental-rotation tasks are included in the SPACE AI model benchmark~\cite{ramakrishnan_does_2024}.

\paragraph{Syllogistic reasoning}
Each item consists of two premises and a conclusion, judged \emph{Valid}, \emph{Invalid}, or \emph{Cannot be determined}. The premises are record-keeping narratives in six everyday scenarios, paraphrasing the quantifiers (\emph{all}, \emph{no}, \emph{some}) with neutral category terms (Appendix~\ref{app:gen:syl}). This paradigm has been studied in humans~\cite{khemlani_theories_2012} and AI models~\cite{eisape-etal-2024-systematic,ozeki-etal-2024-exploring}; designed difficulty uses the mental-model count of \citet{JohnsonLaird1984}.

\paragraph{Letter-String analogies}
Following human and AI model studies by \citet{Lewis2024} and \citet{Webb2022}, items show two worked examples of a transformation and a target sequence to complete as free text. The task builds on the Copycat analogy paradigm~\cite{hofstadter1995copycat}, with difficulty manipulations following \citet{Lewis2024}. Each item states a fictional alphabet order defining the previous and next symbol. We ignored spacing and capitalization when scoring, and reviewed incorrect answers in all groups manually, accepting additional answers that applied a rule consistent with the worked examples, see Appendix~\ref{app:gen:ls}.

\label{sec:method:measures}

\paragraph{Task performance and self-assessment} We scored task performance as the number of correct responses out of $40$, and per task type out of $10$. Participants also assessed their performance at three levels: once before and once after the battery, once after each block, and after every scored item. \autoref{tab:metacog} lists each question and the group it was asked in. From these estimates we derive two measures. Metacognitive accuracy is the signed difference between estimated and achieved score, where positive values indicate overestimation. Metacognitive sensitivity is the area under a participant's type-2 ROC curve (AUROC), which measures how well their item-level confidence separates their correct from their incorrect answers. AUROC is undefined for a participant with no correct or no incorrect answers, and we omit those cases from the sensitivity analysis. We disabled back-navigation throughout, so participants could not revise an estimate after seeing later items.

\paragraph{Additional Questionnaires} Four questionnaire sections followed the battery, covering demographics, frequency of AI use, and ratings of the task set. In \HAI\, participant's additionally described their consultation strategy and rated the AI model's helpfulness, their trust in it, and their frustration (Appendix~\ref{app:questionnaire}).

\begin{table}[h!]
\centering
\caption{Self-assessment questions by level and timepoint, with the response format and the group each question was asked in.}
\label{tab:metacog}
\small
\begin{tabular}{r p{0.62\linewidth} l c c}
\toprule
\multicolumn{2}{l}{Question} & \shortstack{Response\\format} & \shortstack{Human\\alone} & \shortstack{Human\\+AI} \\
\midrule
\multicolumn{5}{l}{\textbf{Global:} Asked once before the first block and repeated after the last block in past tense} \\
\addlinespace
1 & Compared to other participants, rate your general cognitive reasoning ability. & Slider & \checkmark & \checkmark \\
2 & How many of the 40 problems will you solve correctly? & Count & \checkmark & \checkmark \\
3 & Without the AI model, how many of the 40 would you solve correctly? & Count & & \checkmark \\
4 & Compared to other AI systems, rate this AI model's reasoning ability & Slider & & \checkmark \\
5 & On its own, how many of the 40 would the AI model solve correctly? & Count & & \checkmark \\
6 & Compared to other participants, how well will you perform overall? & Slider & \checkmark & \checkmark \\
7 & How difficult do you find cognitive reasoning problems in general? & Likert & \checkmark & \checkmark \\
8 & How difficult are they for the average participant? & Likert & \checkmark & \checkmark \\
\midrule
\multicolumn{5}{l}{\textbf{Block:} Asked after the last item of each of the four blocks} \\
\addlinespace
9 & How many of the 10 [task type] questions do you think you solved correctly? & Count & \checkmark & \checkmark \\
10 & How many would the AI model have solved correctly on its own? & Count & \checkmark & \checkmark \\
11 & Without the AI model, how many would you have solved correctly? & Count & & \checkmark \\
12 & If you had used an AI model, how many would you and it together have solved? & Count & \checkmark & \\
\midrule
\multicolumn{5}{l}{\textbf{Item:} Asked after every scored item} \\
\addlinespace
13 & How confident are you that your response is correct? & Slider & \checkmark & \checkmark \\
\bottomrule
\end{tabular}
\vspace{2pt}
\parbox{\linewidth}{\footnotesize\textit{Note.} Counts are entered as a number, out of 40 for global questions and out of 10 for block questions. Sliders run from 0 to 100, and the confidence slider (question 13) is anchored from \emph{Unsure} to \emph{Certain}. Questions 7 and 8 use a 1--10 Likert scale from very easy to very difficult. In \HAI, questions 1 and 2 refer to ability and performance \textbf{with} the AI model (with / using the AI assistant).}
\end{table}

\subsection{Procedure}
\label{sec:method:procedure}

Participants arrived from Prolific, read the study information and privacy notice, and gave informed consent before entering the proctored study. We informed them of their right to withdraw at any time; consent and proctoring procedures are detailed in Appendix~\ref{app:procedure_detail}.

Participants in \HAI\ saw a tutorial describing the AI model as state-of-the-art but warning that it ``can also make mistakes, so always apply your own judgement''. Participants in \Hum\ saw no equivalent page. Both groups then completed the set-up page, with incentive information, group-specific tips and two attention checks, followed by the global pre-task self-assessment (see \autoref{tab:metacog}, Global). They then completed the four task blocks, each one starting with task instructions and one practice trial, followed by ten task items with a question on confidence after each (see \autoref{tab:metacog}, Item). The block concluded with questions on self-assessment (see \autoref{tab:metacog}, Block). We never cleared the chat history, so the AI model carried the full conversation context across items and blocks. The session closed with the global post-task self-assessment, additional questionnaires, and the leaderboard. A session lasted a median of $77$ minutes.

\subsection{AI Model Benchmarking }
\label{sec:method:benchmark}

To obtain an AI model alone reference, we ran GPT-5.6 Luna at low reasoning effort, the study's setting, over the same $40$ items in 100 chat-parity runs after data collection. The reference is the mean across runs, with the run as the repeated unit on one fixed item set.

The chat-parity benchmark used the participant-facing message text and images without the earlier answer-only system prompt or format restriction. Free-form answers were extracted and scored. Unlike the live study, the benchmark omitted instructions and practice context and reset history at each block; it therefore approximates, but does not reproduce, the study's interaction context. Appendix~\ref{app:harness} documents the prompt construction, answer matching and run bookkeeping.
The primary analysis retains the original recorded scores for the 12 flagged letter-string records. We report the confirmed extraction correction separately as a sensitivity.\footnote{Review of the raw replies identified one complete correct answer missed by extraction (run 92, item \texttt{ls\_028}); crediting it raises the benchmark from $27.71$ to $27.72/40$. The three partial replies remain incorrect under the complete-answer rule. Treating all 12 flagged records as correct gives $27.83/40$; excluding them and rescaling each run's block accuracy gives $27.80/40$, with a Human+AI difference of $0.56$ points, 95\% CI $[-0.36,1.48]$, $p=.235$. Exclusion changes the item set. Unflagged replies have not been audited.}

\subsection{Data Analysis}
\label{sec:method:analysis}
This exploratory study was not preregistered. We analyzed performance, performance estimates, signed errors, and confidence separately by group, using one-sample $t$-tests against reference values, paired $t$-tests for within-participant contrasts, and Welch tests between groups. We report $95\%$ parametric confidence intervals, $d_z$ for paired contrasts, Hedges' $g$ between participant groups, and Cohen's $d$ against the \AI\ benchmark (participants and benchmark runs are the observational units). Holm corrections cover four block comparisons per outcome; within-block reference tests are exploratory and unadjusted.

Exploratory Bayesian equivalence checks used Gaussian mean models in \texttt{brms} \cite{burkner2017brms}, reporting $P(|d|<.30\mid\mathrm{data})$ with a $.95$ criterion and a post-hoc margin. Outcome-specific findings appear in footnotes; response scaling, all equivalence priors, and sensitivity margins are in Appendix~\ref{app:analysis_details}. These assess mean differences, not endpoint mass.

Exploratory component-belief regressions predicted pair estimates from human and AI model estimates, separately by group, with participant and block fixed effects and participant-clustered CR1 covariance. Unconstrained slopes and their difference describe relative associations. Post-task percentiles were tested against $50$ and between groups, with Holm correction across the two measures within each testing family. Pooled achieved ranks, averaging ties, contextualized these judgments rather than correcting score ceilings.

We examined the DKE through within-group lowest-minus-highest quartile contrasts in signed error and estimate--score regression slopes. Disjoint-score checks adapted the separation of ranking and outcome performance \cite{krueger2002unskilled}; split-half instrumental variables adapted the measurement-error correction of \citet{feld2017estimating}. Block regressions used performance on the other $30$ items. Further checks examined curvature and residual spread \cite{gignac2020dunning}, bounded responses, and specified constant-bias/noise simulations. We retained $4{,}000$ participant bootstrap resamples for nonlinear quartile contrasts, IV ratios, and two-stage residual-spread analyses. Appendix~\ref{app:robustness_details} distinguishes sampling intervals from split-sensitivity ranges and null-simulation intervals and documents study-specific adaptations.

For a post-hoc AUROC sensitivity, we restricted correct--incorrect confidence comparisons to the same task block. Blocks were weighted by eligible pairs within each participant, and participants equally within groups; tied confidence received half credit. Blocks without both correct and incorrect answers contribute no pairs, so ceiling-related selection remains. Mean absolute total-score error was reported separately from signed bias.

Item correctness and confidence were predicted by standardized AI model accuracy across $100$ benchmark runs using logistic and linear regressions, respectively, with crossed participant, block, and item random intercepts, separately by group. Wald intervals condition on benchmark estimates. We decomposed overestimation into summed rescaled confidence minus achieved score and the global estimate minus that sum, regressing each component on score. This assumes the Unsure--Certain scale can represent probabilities. We also compared global with summed block estimates.

Finally, total-only and extended bounded-binomial computational models describe achieved scores and estimates jointly. The former uses one achieved and estimated total per person; the extension uses four achieved block scores, four block estimates, and one global estimate, without double-counting achieved totals. Both infer latent skill, group bias, and probit-scale compression; the extension adds block difficulties, person-specific report offsets, and global-report shifts. Section~\ref{sec:results:computational} defines their likelihoods and all priors.

Bayesian fits used four chains of $20{,}000$ iterations, including $4{,}000$ warmup, yielding $64{,}000$ retained draws. We report posterior medians, $95\%$ credible intervals, convergence and Monte Carlo diagnostics. Computational fit checks compare conditional quartile predictions and $2{,}000$ replicated datasets with observed means, spreads, ceiling frequencies, and block/global associations; these are in-sample checks, not held-out validation.
The binomial likelihood includes endpoint probabilities but does not recover unreported estimates beyond the maximum. Smooth computational curves vary latent skill and integrate over person-specific report offsets; their horizontal axis represents expected achieved performance, not observed scores conditioned back to skill.

\section{Results}

\subsection{Human-AI Joint Performance}
\label{sec:res_composite}

We investigate whether the Human-AI pair outperforms both parts (Human and the AI) (\textit{synergy}) or only the Human working alone (\textit{augmentation}), and whether participants know their solo ability, the AI model's ability, and the ability of the pair (\textbf{RQ1}). \textit{Synergy} requires the pair to exceed the better of its two parts, whereas \textit{augmentation} means that the pair exceeds only the human. Throughout, we write \HAI for the score achieved by the Human+AI group, \emph{Human alone} for the score achieved by the Human alone group and \emph{AI model alone} for the AI model's own score, benchmarked on the same 40 items over 100 chat-parity runs\footnote{Contrasts between Human+AI and Human alone (both participant groups) are reported as Hedges' $g$, which corrects the small-sample bias in Cohen's $d$; contrasts against AI model alone are reported as Cohen's $d$.}.

\paragraph{Human+AI team against the AI model alone.} The AI model alone scored $M = 27.71$ ($SD = 2.52$, 95\% CI $[27.21, 28.21]$), and Human+AI achieved $M = 28.36$ ($SD = 5.28$), a difference of $+0.65$ $[-0.27, 1.57]$, Welch's $t(271.70)=1.38$, $p=.168$, $d=0.14$ (100 benchmark runs). $56\%$ of Human+AI participants exceeded the AI model's mean score. This overall comparison does not establish synergy or equivalence\footnote{Exploratory equivalence check using the Gaussian mean model in \texttt{brms}: posterior $d=0.142$, 95\% CrI $[-0.062,0.345]$, and $P(|d|<.30\mid\mathrm{data})=.937$. This falls below the $.95$ criterion, so equivalence is inconclusive. The wider-prior probability was $.932$; priors and standardization are defined in Methods. Retaining original scoring as primary, a Welch-test sensitivity correcting the one confirmed extraction error gave a benchmark mean of $27.72/40$ and a total contrast of $+0.64$, CI $[-0.28,1.56]$, $p=.174$; the block-comparison conclusions were unchanged.}.

\paragraph{Human+AI team against the human alone.} Human+AI scored above Human alone ($M = 21.13$, $SD = 6.67$) by $7.22$ points $[5.99, 8.46]$, $t(351.72) = 11.52$, $p < .001$, $g = 1.20$ (see \autoref{fig:teaser}A). In line with \citet{fernandes_ai_2024}, the group means are consistent with augmentation, with no demonstrated overall synergy.

\begin{figure}[!ht]
  \centering
  \includegraphics[width=0.9\linewidth]{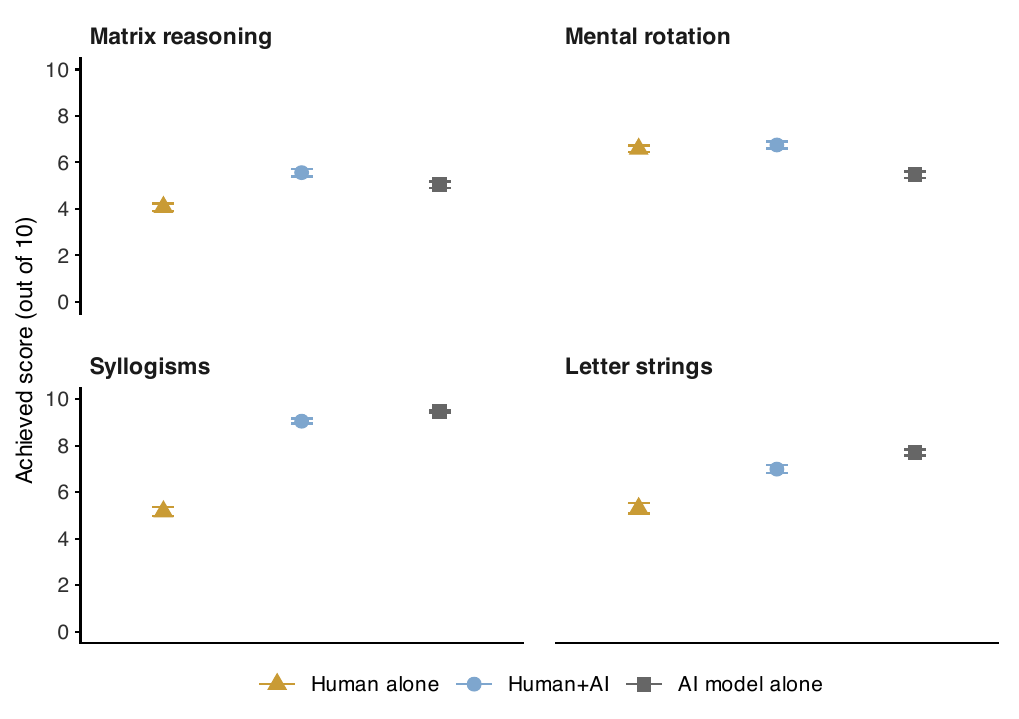}
  \caption{Block-level performance for Human alone (yellow), Human+AI (blue), and AI model alone (grey; 100 chat-parity runs). Points show block means $\pm1$ SEM across participants or AI model runs on the fixed item set. Only matrix reasoning demonstrably exceeds both components. Overall performance is shown in \autoref{fig:teaser}A.}
  \Description{Block mean comparisons for the human, AI model and pair, with one standard error of the mean.}
  \label{fig:performance_combined}
  \label{fig:pairparts}
\end{figure}

\paragraph{Synergy and augmentation by block.} Against the AI model alone, Human+AI scored higher on matrix reasoning, $+0.52$ (95\% CI $[0.11,0.93]$, $t(264.19)=2.48$, $p_{\mathrm{Holm}}=.014$, $d=0.28$), and mental rotation, $+1.28$ ($[0.88,1.67]$, $t(246.73)=6.38$, $p_{\mathrm{Holm}}<.001$, $d=0.74$), but lower on letter strings, $-0.73$ ($[-1.15,-0.30]$, $t(276.85)=-3.34$, $p_{\mathrm{Holm}}=.002$, $d=-0.36$), and syllogisms, $-0.42$ ($[-0.65,-0.20]$, $t(252.01)=-3.69$, $p_{\mathrm{Holm}}<.001$, $d=-0.37$) (see \autoref{fig:pairparts}). Relative to Human alone, Human+AI scored higher by $1.48$ points on matrix reasoning, CI $[1.04,1.93]$, $t(361.81)=6.51$, $p_{\mathrm{Holm}}<.001$, $g=0.68$, supporting task-specific synergy under this benchmark protocol. On mental rotation, Human alone also exceeded AI model alone ($6.58$ versus $5.47$), and the Human+AI--Human alone difference was uncertain ($+0.17$, CI $[-0.21,0.55]$, $t(363.78)=0.88$, $p_{\mathrm{Holm}}=.381$, $g=0.09$). Human+AI gains over Human alone were $+3.88$ on syllogisms (CI $[3.46,4.31]$, $t(285.88)=18.12$, $p_{\mathrm{Holm}}<.001$, $g=1.87$) and $+1.68$ on letter strings (CI $[1.12,2.25]$, $t(341.01)=5.85$, $p_{\mathrm{Holm}}<.001$, $g=0.61$).\footnote{For Human+AI versus Human alone on mental rotation, posterior $d=0.090$, 95\% CrI $[-0.113,0.294]$, with $P(|d|<.30\mid\mathrm{data})=.978$ (wider prior: $.977$), supporting practical equivalence at the post-hoc $.30$ margin. The tighter $.20$ margin remained inconclusive ($P=.850$).}

\begin{table*}[t]
\centering
\caption{Whole-battery descriptives by group and performance quartile (Q). Cells are $M$ ($SD$). The whole-battery counterfactual and whole-battery estimate of AI model alone were asked only in Human+AI; both groups provided block-level AI model estimates. AUROC uses $n=178$ in Human+AI ($44$ in Q4), with all other participants included. Percentile judgments use a 0--100 scale.}
\label{tab:desc_robin}
\small
\setlength{\tabcolsep}{4pt}
\begin{tabularx}{\linewidth}{@{}Xrrrrr@{}}
\toprule
\multicolumn{6}{l}{\textbf{Human alone}} \\
Measure & Full & Q1 & Q2 & Q3 & Q4 \\
\midrule
$n$ & 187 & 47 & 47 & 46 & 47 \\
Performance (0--40) & 21.13 (6.67) & 12.53 (2.49) & 18.83 (1.62) & 23.59 (1.38) & 29.64 (2.85) \\
Estimate, post-task & 23.19 (7.68) & 20.00 (7.77) & 20.85 (8.33) & 23.67 (6.81) & 28.23 (4.61) \\
Estimate, pre-task & 26.77 (6.09) & 25.04 (5.43) & 25.62 (7.51) & 27.78 (5.12) & 28.66 (5.43) \\
Metacog. accuracy (est.\ $-$ perf.) & 2.05 (7.91) & 7.47 (7.97) & 2.02 (8.64) & 0.09 (6.81) & -1.40 (4.82) \\
Estimate without AI (counterfact.) & -- & -- & -- & -- & -- \\
Estimate of the AI model alone & -- & -- & -- & -- & -- \\
Metacog. sensitivity (AUROC) & 0.65 (0.12) & 0.65 (0.10) & 0.65 (0.10) & 0.65 (0.14) & 0.65 (0.14) \\
Mean item confidence (0--1) & 0.66 (0.18) & 0.56 (0.20) & 0.65 (0.19) & 0.70 (0.14) & 0.72 (0.15) \\
Perceived difficulty, self & 7.34 (1.89) & 7.83 (2.07) & 7.94 (1.77) & 7.04 (1.66) & 6.55 (1.74) \\
Perceived difficulty, avg.\ participant & 7.19 (1.67) & 7.43 (1.74) & 7.06 (1.82) & 6.98 (1.58) & 7.30 (1.53) \\
Performance percentile & 50.41 (24.19) & 42.64 (25.50) & 43.94 (24.94) & 54.07 (19.08) & 61.09 (22.44) \\
Reasoning-ability percentile & 50.90 (23.86) & 43.28 (24.59) & 45.11 (24.91) & 55.13 (20.30) & 60.17 (21.76) \\
\midrule
\multicolumn{6}{l}{\textbf{Human+AI}} \\
Measure & Full & Q1 & Q2 & Q3 & Q4 \\
\midrule
$n$ & 179 & 45 & 45 & 44 & 45 \\
Performance (0--40) & 28.36 (5.28) & 21.89 (3.50) & 26.53 (1.06) & 30.00 (1.29) & 35.04 (1.73) \\
Estimate, post-task & 30.37 (5.88) & 31.27 (6.72) & 28.33 (6.45) & 31.07 (4.78) & 30.84 (5.01) \\
Estimate, pre-task & 32.15 (6.01) & 32.07 (6.58) & 30.62 (6.58) & 33.34 (5.03) & 32.58 (5.55) \\
Metacog. accuracy (est.\ $-$ perf.) & 2.02 (7.62) & 9.38 (6.86) & 1.80 (6.59) & 1.07 (4.54) & -4.20 (5.35) \\
Estimate without AI (counterfact.) & 21.37 (8.27) & 20.71 (8.87) & 18.78 (9.11) & 22.82 (7.23) & 23.20 (7.15) \\
Estimate of the AI model alone & 29.34 (7.59) & 30.96 (8.30) & 29.47 (9.03) & 30.20 (5.70) & 26.73 (6.34) \\
Metacog. sensitivity (AUROC) & 0.56 (0.12) & 0.55 (0.12) & 0.58 (0.10) & 0.56 (0.12) & 0.57 (0.13) \\
Mean item confidence (0--1) & 0.78 (0.15) & 0.79 (0.14) & 0.71 (0.15) & 0.80 (0.15) & 0.81 (0.13) \\
Perceived difficulty, self & 7.05 (1.99) & 7.13 (2.16) & 7.02 (1.88) & 7.00 (2.25) & 7.04 (1.68) \\
Perceived difficulty, avg.\ participant & 7.19 (1.84) & 7.16 (2.13) & 7.04 (1.81) & 7.25 (1.73) & 7.31 (1.72) \\
Performance percentile & 59.95 (23.30) & 63.87 (21.12) & 51.89 (23.41) & 62.23 (25.18) & 61.87 (22.18) \\
Reasoning-ability percentile & 63.37 (22.46) & 65.80 (23.68) & 56.04 (22.64) & 66.09 (22.72) & 65.62 (19.74) \\
\bottomrule
\end{tabularx}
\end{table*}

With at least one message required per item, $4\%$ of Human+AI participants never sent more than one prompt on any item, whereas $18\%$ sent more than five on at least one item (Appendix Table~\ref{tab:prompts_robin}).

\subsubsection{Knowing the self, the AI model, and the pair}

The second part of RQ1 compares participants' beliefs about their solo performance, the AI model's performance, and the pair's performance. Human alone participants' self-estimates and Human+AI participants' pair estimates were compared with their own achieved scores. AI model estimates were compared with the independent benchmark on the same 40 items. Hypothetical pair estimates in Human alone and hypothetical solo estimates in Human+AI were compared with the other group's block means, not with observed individual counterfactual scores.

To test whether Human+AI participants' estimates of AI model performance were accurate, we compared their estimate for each block with the AI model's benchmarked score in the same block. 
At the group mean, their estimate (7.47 out of 10 per block) exceeded the AI model's measured score (6.93) by 0.55 points. 

\begin{figure}[!ht]
  \centering
  \includegraphics[width=0.9\linewidth]{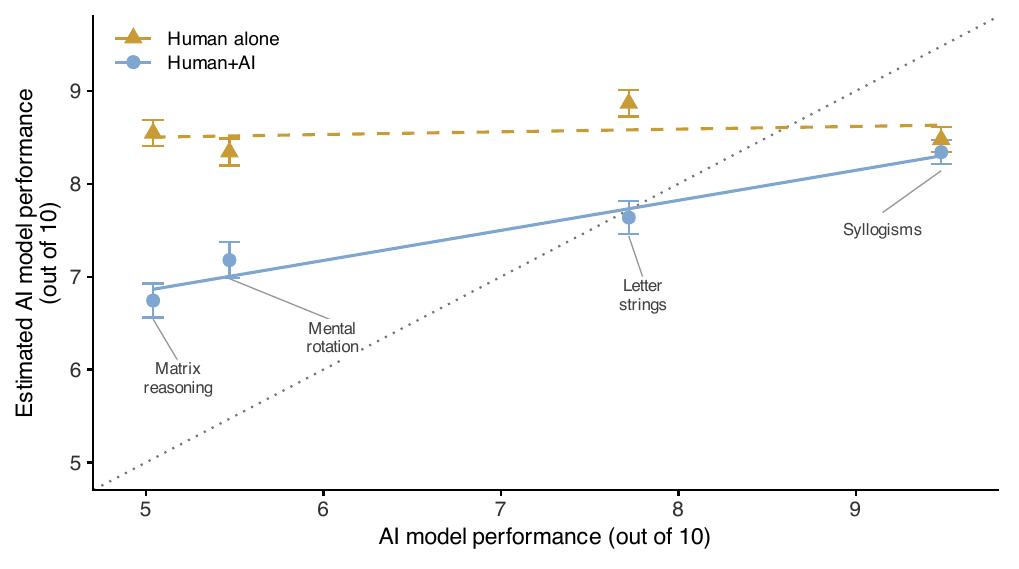}
    \caption{Comparison of the estimated and actual score of the AI model alone per block (out of 10) for Human alone (yellow) and Human+AI (blue). Points represent group means $\pm1$ SEM, and lines represent linear fits. The dotted line indicates perfect knowledge of the AI model's score. Both groups' estimates vary less across blocks than the AI model's actual score.}

    \Description{Believed versus actual AI model performance per block for both groups.}
  \label{fig:knowmodel}
\end{figure}

To see whether their estimates followed the AI model's performance across blocks, we regressed each participant's four estimates on the AI model's benchmarked score in the same blocks and report the mean of the resulting coefficients with a 95\% confidence interval.
Across the four blocks, AI model scores ranged from $5.04$ to $9.48$. Human+AI estimates followed this variation with a coefficient of $0.324$ (95\% CI $[0.241,0.407]$), where 1 would indicate perfect tracking, and a mean absolute error of $2.15$ points per block (\autoref{fig:knowmodel}). The Human alone slope was $0.029$ ($[-0.027,0.085]$), with mean overestimation of $1.63$ points per block. Therefore, human+AI estimates only weakly tracked the performance of AI models across blocks; a modest signed mean error did not imply accurate block-level judgments.

Human+AI participants' estimates of solo performance were $0.47$ points per block above the Human alone group's performance mean. Their AI model estimates exceeded the benchmark by $0.55$, and their pair estimates exceeded their own achieved scores by $0.71$. Human alone participants overestimated their own scores by $0.41$ and the AI model benchmark by $1.63$. Their hypothetical pair estimates were $1.89$ points higher than the Human+AI group average. The pair had the largest mean signed difference from its respective reference in both groups. Because these references combine personal scores and group means, this comparison does not establish that the pair is intrinsically harder to estimate.

To compare expected and observed collaboration benefits, we first examine Human alone participants' hypothetical pair estimates alongside their human alone and AI model estimates. We subtracted the higher of their two component estimates (i.e., their own performance or the AI model's) from their pair estimate. Human alone participants expected the pair to exceed the higher of its parts by $0.152$ points per block (95\% CI $[0.019, 0.286]$, $t(186)=2.25$, $p=.025$, $d_z=0.16$), testing each participant's mean across the four blocks, whereas the pair scored $0.11$ points below the higher measured part when averaged across blocks, a descriptive comparison between within-person beliefs and measured group means. In line with \citet{fernandes_ai_2024}, estimates were inflated in both groups. These summaries use different reference levels and do not estimate individual counterfactual gains.

\paragraph{Combining beliefs about the two parts.}
To see how participants combined their beliefs about the two components, we regressed their estimate of the pair on their estimate of themselves and their estimate of the AI model, separately for each group, in an exploratory analysis.  Participant-clustered intervals account for repeated reports. Among Human alone participants, the estimate of the pair followed their estimate of the AI model more closely ($b=0.409$, 95\% CI $[0.272,0.546]$) than their estimates ($b=0.141$, $[0.080,0.202]$), a difference of $0.268$ ($[0.107,0.429]$, $t(186)=3.29$, $p=.001$).
Among Human+AI participants, who estimated the pair they had actually worked as, the estimate of the pair followed both their estimate of themselves without the AI model  ($b=0.364$, $[0.279,0.449]$) and their estimate of the AI model ($b=0.284$, $[0.217,0.351]$). The difference between the two, ($-0.080$, $[-0.202,0.042]$, $t(178)=-1.30$, $p=.197$), could not be distinguished from zero.

Pair estimates (Human+AI) equaled the higher of their two estimates in 63.1\% of Human alone reports and 49.4\% of Human+AI reports, and exceeded both of them in 24.6\% and 22.1\%, respectively. Thus, a weighted average of the two estimates cannot account for every judgment of the pair. The stronger association with the AI model estimate in the Human alone group remained when reports with any estimate at 10 were excluded (coefficient difference $0.315$, $[0.067,0.563]$, $t(120)=2.51$, $p=.013$; 287 reports from 121 participants). Note that these coefficients are in points of the pair estimate per point of the estimate of the human or the AI model, and not percentage weights.

\paragraph{Post-task percentile judgments.}
We also examined relative self-assessment, which does not express error as an estimated count minus a score near the maximum of 40. Participants rated their overall performance and general reasoning ability on a 0--100 percentile scale. In the Human+AI group, the ability question concerned reasoning when using the AI model (see \autoref{tab:metacog}), so this is not a comparison of perceived solo ability. We tested each group against 50 and compared groups with Welch's tests. 

For overall performance, Human alone reported $M=50.41$ ($SD=24.19$), a deviation of $+0.41$ percentile points from 50, CI $[-3.08,3.90]$, $t(186)=0.23$, $p=.816$, $d=0.02$. Human+AI reported $M=59.95$ ($SD=23.30$), a deviation of $+9.95$, CI $[6.51,13.39]$, $t(178)=5.71$, $p<.001$, $d=0.43$. The group difference was $9.54$ percentile points, CI $[4.66,14.42]$, Welch's $t(363.98)=3.84$, $p<.001$, $g=0.40$.

For general reasoning ability, Human alone reported $M=50.90$ ($SD=23.86$), a deviation of $+0.90$, CI $[-2.54,4.34]$, $t(186)=0.51$, $p=.607$, $d=0.04$. Human+AI reported $M=63.37$ ($SD=22.46$), a deviation of $+13.37$, CI $[10.06,16.69]$, $t(178)=7.97$, $p<.001$, $d=0.60$. The group difference was $12.48$ percentile points, CI $[7.71,17.24]$, Welch's $t(363.90)=5.15$, $p<.001$, $g=0.54$. The Human+AI midpoint tests and both between-group tests remain significant after Holm correction across the two percentile measures within each test family.\footnote{For Human alone versus the midpoint of 50, posterior $d=0.017$, 95\% CrI $[-0.126,0.160]$, for performance placement, and $d=0.037$, $[-0.107,0.182]$, for ability placement. Both had $P(|d|<.30\mid\mathrm{data})>.999$, supporting equivalence to the scale midpoint, not to actual rank. Both also exceeded $.95$ at the $.20$ margin; conclusions were unchanged with the wider prior.} These findings establish higher relative self-placement in Human+AI.%

\subsection{Metacognitive accuracy and metacognitive sensitivity}
\label{sec:res_metacog}

RQ2 asks whether AI assistance changes metacognitive accuracy and metacognitive sensitivity. Metacognitive accuracy refers to how closely participants' estimated performance aligns with their actual performance, and metacognitive sensitivity to how well their confidence on each item distinguishes correct from incorrect answers. 

\subsubsection{Metacognitive accuracy}

Investigating metacognitive accuracy on the whole battery for each group, we find that participants overestimated their performance in both groups, which aligns with \citet{fernandes_ai_2024}. 
In the Human+AI group, participants overestimated their performance by $M = +2.02$ ($SD = 7.62$) points, 95\% CI $[0.89,3.14]$, $t(178) = 3.54$, $p < .001$, $d = 0.26$, and in the Human alone group $M = +2.05$ ($SD = 7.91$), CI $[0.91,3.19]$, $t(186) = 3.55$, $p < .001$, $d = 0.26$ (see also \autoref{tab:desc_robin}). The between-group difference was of $-0.04$ $[-1.63, +1.56]$, $t(363.98) = -0.05$, Welch $p = .964$, $g = 0.00$. The exploratory equivalence checks supported small group differences in signed and absolute mean error. Mean absolute errors were $6.26$ points in Human alone and $6.15$ in Human+AI (difference $-0.11$, CI $[-1.15,0.94]$, $t(363.87)=-0.20$, $p=.842$, $g=-0.02$): small signed group means therefore do not establish accurate individual self-assessment.\footnote{The \texttt{brms} equivalence checks gave posterior $d=-0.004$, 95\% CrI $[-0.206,0.198]$, for signed error ($P(|d|<.30\mid\mathrm{data})=.996$), and $d=-0.019$, $[-0.223,0.186]$, for absolute error ($P=.995$). Both support practical equivalence of group means at the post-hoc $.30$ margin, with the same conclusions under the wider prior. Neither supported equivalence at $.20$; practical equivalence does not mean accurate individual self-assessment.}

Comparing estimates before and after the battery, we find that both groups lowered their estimates after completing it. Human alone participants' estimates decreased from $26.77$ ($SD = 6.09$) to $23.19$ ($SD = 7.68$), a difference of $-3.58$  (95\% CI
$[-4.58, -2.59]$, $t(186) = -7.08$, $p < .001$, $d_z = -0.52$). Human+AI participants' estimates decreased from $32.15$ ($SD = 6.01$) to $30.37$ ($SD = 5.88$), a difference of $-1.77$ $[-2.75, -0.79]$, $t(178) = -3.57$, $p < .001$, $d_z = -0.27$. Therefore, the Human alone group corrected its estimate by $1.81$ points more than the Human+AI group, $[+0.42, +3.20]$, Welch's $t(364)=2.56$, $p=.011$, $g=0.27$.

\subsubsection{Metacognitive sensitivity}
To see whether participants track information on each item, we turn to metacognitive sensitivity, which we estimate from the confidence ratings given after each answer. We conducted a ROC analysis for each participant and computed the area under the ROC curve (AUROC), which quantifies how well a participant's confidence ratings differentiate between correct and incorrect responses; an AUROC of $.5$ indicates no better-than-chance discrimination and $1$ perfect discrimination. The mean AUROC was $.649$ ($SD = 0.122$) in the Human alone group and $.564$ ($SD = 0.120$) in the Human+AI group. Both exceeded $.5$ (Human alone: deviation $0.149$, 95\% CI $[0.131,0.166]$, $t(186)=16.75$, $d=1.22$; Human+AI: deviation $0.064$, CI $[0.046,0.081]$, $t(177)=7.08$, $d=0.53$; both $p<.001$). One Human+AI participant had no incorrect answers and hence no defined AUROC; its valid group sample was $n=178$. The Human+AI group was thus only slightly above chance, and the difference between the groups was substantial, $-0.085$ $[-0.110,-0.060]$, $t(362.55) = -6.75$, $p < .001$, $g = -0.71$. The confidence gap (i.e., mean confidence on correct items minus mean confidence on incorrect items, on a $0$--$1$ scale) decreased from $0.12$ ($SD = 0.11$) in the Human alone group to $0.04$ ($SD = 0.08$) in the Human+AI group, $-0.08$ $[-0.10, -0.06]$, $t(351.08)=-7.71$, $p < .001$, $g = -0.80$. Mean item confidence, in turn, increased from $0.66$ ($SD = 0.18$) to $0.78$ ($SD = 0.15$), $+0.12$ $[+0.09, +0.16]$, $t(356.22)=7.17$, $p < .001$, $g = 0.74$. Human+AI participants were more confident but showed lower discrimination across the pooled battery. Sensitivity measures can depend on task performance \cite{rahnev2025comprehensive}; the group contrast should not be interpreted as a performance-independent measure of metacognitive efficiency.

As a post-hoc sensitivity check, we restricted correct--incorrect confidence comparisons to the same task block, weighting eligible blocks by their number of such pairs within each participant. Within-block AUROC was $.590$ in Human alone and $.568$ in Human+AI; the difference was $-.022$, 95\% CI $[-.047,.004]$, Welch's $t(362.03)=-1.69$, $p=.091$, $g=-0.18$. Approximately 80\% of pairs in the pooled AUROC crossed task blocks. The smaller within-block contrast suggests an important contribution from cross-task confidence ordering, but does not establish equivalent within-task discrimination. Perfect block scores contribute no within-block correct--incorrect pairs, so this check does not eliminate ceiling-related selection.\footnote{For within-block AUROC, posterior $d=-0.175$, 95\% CrI $[-0.380,0.029]$, with $P(|d|<.30\mid\mathrm{data})=.884$ (wider prior: $.881$). Equivalence remains inconclusive; the nonsignificant group comparison should not be read as equal discrimination.}

\begin{table*}[t]
\centering
\caption{Performance and metacognition by block. Cells are $M$ ($SD$), with scores and estimates out of 10. Accuracy is signed estimation error (estimate minus performance); confidence is on a 0--1 scale. AI model alone was benchmarked over 100 chat-parity runs. Valid AUROC counts for Human+AI/Human alone are 172/177 (matrices), 165/174 (rotation), 93/172 (syllogisms), and 145/151 (letter strings); other measures use 179/187.}
\label{tab:blocks_robin}
\footnotesize
\setlength{\tabcolsep}{3pt}
\resizebox{\textwidth}{!}{%
\begin{tabular}{l r rrrrr rrrrr}
\toprule
& AI model & \multicolumn{5}{c}{Human+AI} & \multicolumn{5}{c}{Human alone} \\
\cmidrule(lr){3-7}\cmidrule(lr){8-12}
Block & alone & Perf. & Est. & Accuracy & AUROC & Conf. & Perf. & Est. & Accuracy & AUROC & Conf. \\
\midrule
Matrix reasoning & 5.04 (1.42) & 5.56 (2.05) & 8.03 (1.73) & 2.47 (2.37) & 0.57 (0.20) & 0.79 (0.16) & 4.07 (2.31) & 5.87 (2.20) & 1.80 (2.37) & 0.60 (0.21) & 0.66 (0.21) \\
Mental rotation & 5.47 (1.45) & 6.75 (1.85) & 8.30 (1.73) & 1.55 (2.49) & 0.54 (0.21) & 0.85 (0.14) & 6.58 (1.88) & 6.40 (2.09) & -0.18 (2.45) & 0.56 (0.25) & 0.71 (0.18) \\
Syllogisms & 9.48 (0.52) & 9.06 (1.37) & 7.50 (2.24) & -1.55 (2.31) & 0.54 (0.30) & 0.74 (0.20) & 5.17 (2.58) & 4.56 (2.30) & -0.61 (2.35) & 0.51 (0.23) & 0.59 (0.22) \\
Letter strings & 7.72 (1.32) & 6.99 (2.31) & 7.38 (2.56) & 0.39 (2.65) & 0.61 (0.22) & 0.73 (0.23) & 5.31 (3.15) & 5.93 (3.02) & 0.62 (2.49) & 0.71 (0.24) & 0.66 (0.25) \\
\bottomrule
\end{tabular}}
\end{table*}

Estimated and actual score correlated at $r=.40$ in Human alone and $r=.07$ in Human+AI (Appendix Table~\ref{tab:cor_robin}). The Human+AI correlation was close to zero, but its uncertainty does not establish that estimates contain no information about scores.

\subsubsection{Accuracy and sensitivity by block}

Our data shows that overestimation tracked the difficulty of the block (see 
\autoref{tab:blocks_robin}), alongside the AI model's benchmarked score on the same items. 
Human+AI participants overestimated their performance by $2.47$ points on matrix reasoning, the block on which they scored lowest, and underestimated it by $1.55$ points on syllogisms, the block on which they scored highest ($9.06$ out of 10). The Human alone group also overestimated most on matrix reasoning, its hardest block ($+1.80$), and underestimated on syllogisms ($-0.61$). This pattern aligns with the ``hard-easy effect'' \cite{moore2008trouble}, where people overestimate their score on difficult tasks and underestimate it on easy ones.

\begin{figure}[!ht]
  \centering
\includegraphics[width=0.9\linewidth]{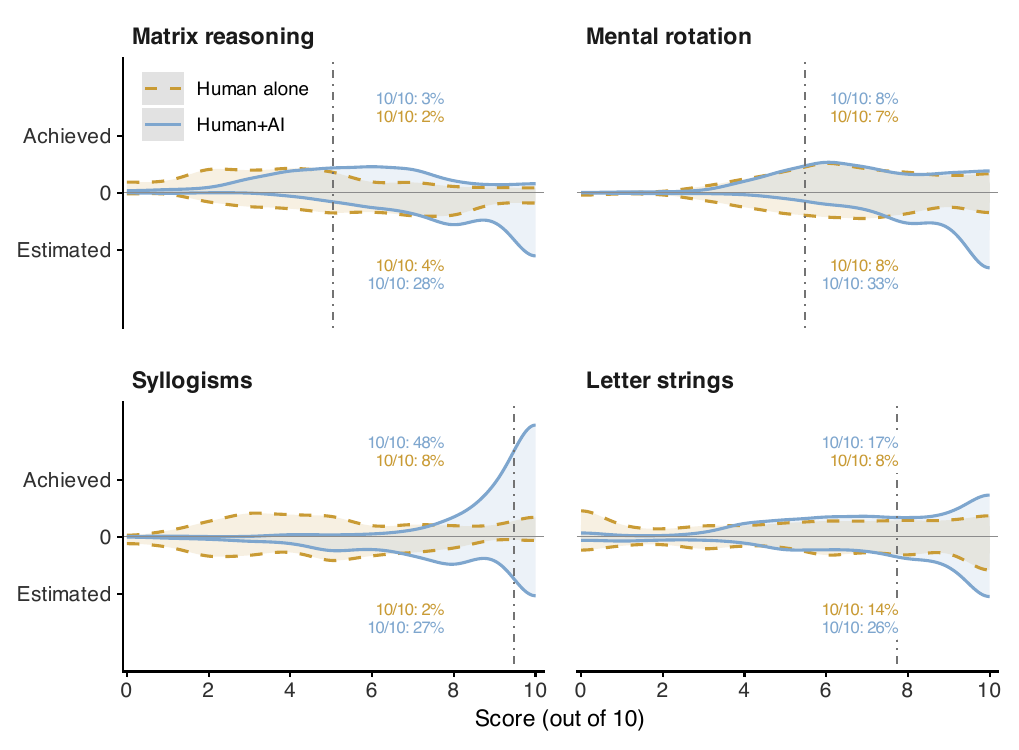}
\caption{Mirrored distributions of achieved scores (above zero) and self-estimated scores (below zero), out of 10 by block. Human alone is yellow and Human+AI blue. Each half has unit area and uses the same boundary-reflected smoothing bandwidth. Labels report the observed percentage at 10/10; smoothing does not represent a continuous or uncensored response scale. Grey dot-dash lines mark mean AI model performance.}
\Description{Achieved and estimated score densities mirrored on a common score axis for each task, with ceiling percentages.}
\label{fig:blockdens}
\label{fig:blockestimates}
\end{figure}

Descriptively, metacognitive sensitivity was lower with AI assistance in three of the four blocks (letter strings: $.61$ vs.\ $.71$; mental rotation: $.54$ vs.\ $.56$; matrix reasoning: $.57$ vs.\ $.60$), but not on syllogisms ($.54$ vs.\ $.51$), where both values are close to chance. Note that Human+AI participants answered almost every syllogism correctly, so that few incorrect items remain from which to compute an AUC. On this block their scores are at the ceiling of the scale while their self-estimates are not, and the reverse holds for matrix reasoning (see \autoref{fig:blockdens}).

\subsection{The Dunning-Kruger Effect (RQ3)}
\label{sec:res_dke}
RQ3 asks whether the DKE survives controls for regression to the mean and measurement error, and how the score--overestimation association is distributed between item-confidence error and aggregation.

\subsubsection{The quartile pattern}

The DKE describes a pattern in which lower performers overestimate their performance while higher performers underestimate it \cite{kruger1999unskilled}.
To test whether it occurs in our data, we split each
group into performance quartiles and compared estimated performance with actual performance within each quartile (\autoref{fig:teaser}B). Comparing the lowest and the highest quartile for each group, we find that the lowest quartile overestimated its performance (i.e., estimated minus actual performance) relatively more when compared to the best-performing quartile (Human+AI: Q1 $M = 9.38$, 95\% CI $[7.32,11.44]$, $t(44) = 9.17$, $p < .001$, $d = 1.37$, Q4 $M = -4.20$, CI $[-5.81,-2.59]$, $t(44) = -5.26$, $p < .001$, $d = -0.78$, Q1$-$Q4 $= 13.58$, CI $[11.00,16.16]$, $t(83.11) = 10.47$, $p < .001$, $d = 2.21$; Human alone: Q1 $M = 7.47$, CI $[5.13,9.81]$, $t(46) = 6.42$, $p < .001$, $d = 0.94$, Q4 $M = -1.40$, CI $[-2.82,0.01]$, $t(46) = -2.00$, $p = .052$, $d = -0.29$, Q1$-$Q4 $= 8.87$, CI $[6.17,11.58]$, $t(75.65) = 6.53$, $p < .001$, $d = 1.35$). 

The difference between the two quartile contrasts was $4.71$ points (participant-bootstrap 95\% CI $[0.05, 7.98]$, with quartiles re-formed in each of 4,000 resamples), so that the observed DKE contrast was larger in the Human+AI group.\footnote{When separate item halves determined ranking and the score used to calculate overestimation, the lowest-to-highest-quartile contrast was $2.38$ points larger in Human+AI than in Human alone (participant-bootstrap 95\% CI $[-2.14,7.22]$). Splitting the battery reduces the information available for ranking and scoring, which can reduce precision. The point estimate was positive, but the uncertainty was too large to resolve the direction of the controlled group difference (\autoref{app:robustness_details}).}

Note that the highest Human alone quartile could not be distinguished from 0, so this test does not establish either underestimation or its absence in that quartile. The Human+AI group thus showed evidence of overestimation in its lowest and underestimation in its highest quartile; in Human alone, only the lowest-quartile overestimation was established.

However, both this pattern and its difference between groups can contain regression-to-the-mean artifacts \cite{krueger2002unskilled, gignac2020dunning}. In \autoref{sec:robust_robin}, we test specified statistical accounts and assess sensitivity to measurement error.

\begin{figure*}[t]
  \centering
  \includegraphics[width=0.9\linewidth]{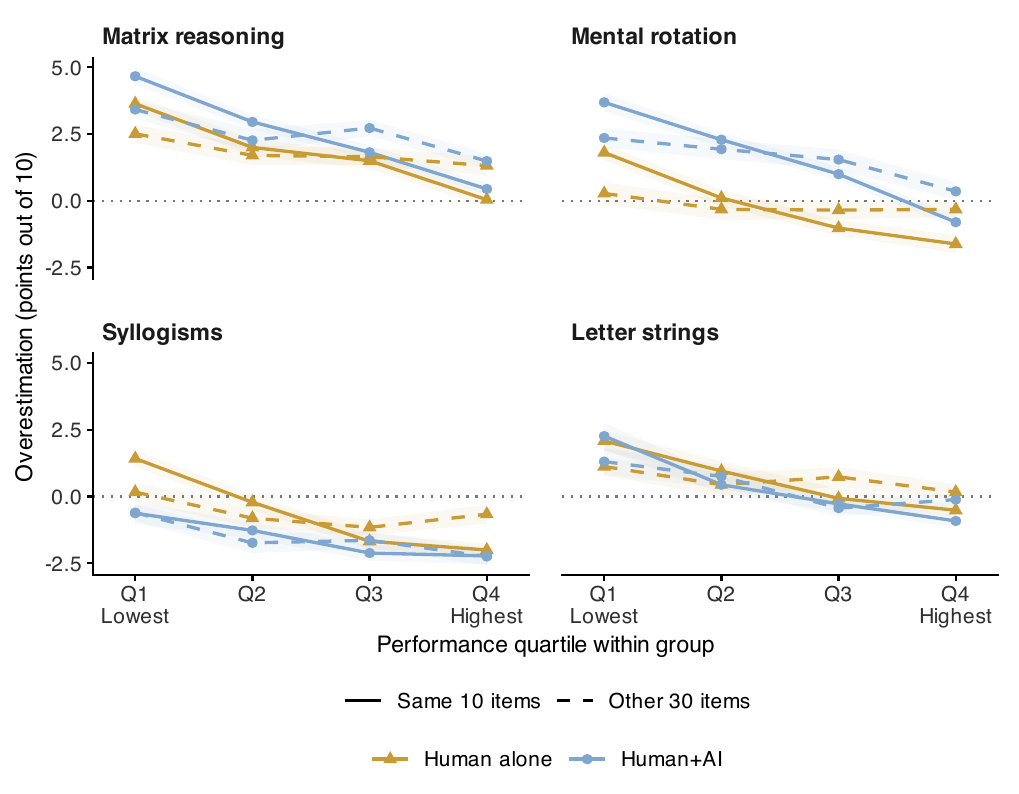}
  \caption{Overestimation (points out of 10) by performance quartile within each block for Human alone (yellow) and Human+AI (blue). Solid lines rank on the same ten items and dashed lines on the other 30 items, separating the ranking from the block score. Shaded bands indicate $\pm1$ SEM around the quartile means. The overall decrease is smaller with the disjoint ranking.}
  \Description{Overestimation by quartile for each block, with and without the regression control.}
  \label{fig:dkeblocks}
\end{figure*}

To see whether the DKE pattern holds within each block, we ranked participants within each block on the block's own ten items and find that overestimation decreased from the lowest to the highest quartile in all four blocks (see \autoref{fig:dkeblocks}, solid lines). Note that this decrease is partly due to regression to the mean (the same ten items determine a participant's quartile and enter their block score). When participants are ranked on the remaining 30 items instead, the decrease is descriptively smaller but does not disappear (dashed lines).

To quantify the same comparison within each block, we fitted a linear regression (ordinary least squares) of block-level overestimation on ability, measured as the standardized score on the other 30 items, separately for each block and each group (i.e., eight models). This adapts the different-test separation principle of \citet{krueger2002unskilled} (Appendix~\ref{app:robustness_details}).
In the Human alone group, the estimates ranged from $-0.443$ (95\% CI $[-0.781, -0.105]$) on matrix reasoning to $-0.259$ $[-0.612, +0.095]$ on mental rotation, and in the Human+AI group from $-0.628$ $[-0.986, -0.270]$ on mental rotation to $-0.508$ $[-0.842, -0.175]$ on syllogisms. Note that these estimates are in points of block overestimation per standard deviation of ability and should thus not be compared to the quartile differences above. Six of the eight unadjusted intervals exclude 0; the two remaining intervals (mental rotation and letter strings, both Human alone) include 0. Descriptively, the decrease was larger in the Human+AI group than in the Human alone group in each block. With Holm correction across four block slopes within each group, all four Human+AI slopes and the Human alone matrix-reasoning slope remained significant. The between-group comparison of these block slopes is descriptive.

\subsubsection{Controls for regression to the mean and measurement error} \label{sec:robust_robin}

The first part of RQ3 investigated whether the DKE survives controls for regression to the mean and measurement error \cite{krueger2002unskilled}. In the strict disjoint-score analysis, odd-numbered items determined quartiles and the outcome was the estimate minus twice the even-item score, so no item contributed to both ranking and the subtracted score. The Q1--Q4 overestimation contrast was $8.87$ points in Human+AI and $6.49$ in Human alone, compared with uncontrolled contrasts of $13.58$ and $8.87$. Preserving shared syllogism scenarios gave the same qualitative conclusion (Appendix~\ref{app:robustness_details}).

To correct estimate--score tracking for measurement error, we used a split-half instrumental variable \cite{feld2017estimating}. Score reliability was $.862$ in Human alone and $.806$ in Human+AI. The corrected tracking slopes were $0.538$ (95\% CI $[0.355,0.722]$) in Human alone and $0.071$ ($[-0.161,0.301]$) in Human+AI, compared with uncorrected slopes of $0.461$ and $0.077$. Both corrected intervals exclude perfect tracking ($1$); the Human+AI interval includes $0$. This interpretation depends on the split-half instrumental-variable assumptions.

Our study-specific constant-bias, constant-noise simulation, motivated by statistical-artifact simulations \cite{gignac2020dunning}, reproduced approximately $29\%$ of the observed Human+AI quartile contrast and $40\%$ of the Human alone contrast; none of 1,500 simulations reached the observed contrast in either group. A separate rounded, censored-normal check of a linear, constant-noise report distribution gave lower-tail $p=.009$ in Human+AI and $p<.001$ in Human alone for the residual-spread statistic, adapting the Glejser diagnostic of \citet{gignac2020dunning}. These results reject those specific accounts, not every statistical explanation. The bounded-null check does not integrate uncertainty in its fitted parameters or observed ability scores. Full procedures and sensitivities can be found in Appendix~\ref{app:robustness_details}.

\subsubsection{Computational models of score self-assessment}
\label{sec:results:computational}

As a descriptive extension of the preceding analyses, we fitted two versions of the bias-and-noise computational account used by \citet{fernandes_ai_2024}. The total-only specification describes one achieved total and one estimated total per participant. The extended specification describes the four achieved block scores, four block estimates, and global estimate jointly. The achieved total is not entered again in the extended likelihood because it is already the sum of the block scores.

Let $i$ index participants, $j$ blocks and $k$ groups. Latent skill has prior $\theta_i\sim\mathcal N(0,2)$; $\Phi_{\mathrm{approx}}$ denotes Stan's approximation to the standard normal CDF:
\begingroup
\begin{align}
y^{\mathrm{obj}}_{ij}&\sim\mathrm{Binomial}\!\left(10,\Phi_{\mathrm{approx}}(\theta_i-d_{kj})\right),\\
y^{\mathrm{per}}_{ij}&\sim\mathrm{Binomial}\!\left(10,\Phi_{\mathrm{approx}}\!\left(\frac{\theta_i-d_{kj}+b_k+u_i}{\sigma_k}\right)\right),\\
y^{\mathrm{glob}}_i&\sim\mathrm{Binomial}\!\left(40,\Phi_{\mathrm{approx}}\!\left(\frac{\theta_i+b_k+u_i+c_k}{\sigma_k}\right)\right).
\end{align}
\endgroup
Here $b_k$ shifts perceived skill, $\sigma_k$ scales the probit predictor, $u_i\sim\mathcal N(0,\tau_k)$ is a shared person-specific report offset, and $c_k$ shifts the global report. Difficulties are centered within group, $d_{kj}=d^{\mathrm{raw}}_{kj}-\overline d^{\mathrm{raw}}_k$, with raw difficulties $\mathcal N(0,1)$. Priors are $b_k,c_k\sim\mathcal N(0,2)$, $\sigma_k\sim\mathrm{LogNormal}(0,2)$, and $\tau_k\sim\mathcal N^+(0,1)$; normal-distribution second arguments denote SDs. Skill is inferred jointly from achieved scores and estimates, not externally measured.

The bounded binomial likelihood includes the endpoint probability at 10 or 40. This corrects the earlier ceiling implementation but does not estimate unobserved reports above the maximum. There were 255 block estimates at 10 (17.42\% of 1,464) and seven global estimates at 40 (1.91\% of 366). The total-only comparator uses the same priors for $\theta$, $b$, and $\sigma$, with probabilities $\Phi_{\mathrm{approx}}(\theta_i)$ for achieved score and $\Phi_{\mathrm{approx}}((\theta_i+b_k)/\sigma_k)$ for estimated score.

Each specification was rerun with four chains of 20,000 iterations: 4,000 warmup and 16,000 sampling, yielding 64,000 post-warmup draws per specification.\footnote{Completed WAMBS checks covered 50 fits \cite{depaoli2017wambs}. Doubling sampling and changing priors left the reviewed score-level and equivalence conclusions unchanged; correlated latent parameters shifted together. Some Monte Carlo precision targets and predictive ceiling fit remained imperfect. Appendix~\ref{app:wambs_review} reports priors, diagnostics, and comparisons.}

The extended specification had maximum $\hat R=1.0004$, minimum bulk ESS $9213$, and minimum tail ESS $15018$. There were 0 divergences and 0 maximum-treedepth hits; chain E-BFMI ranged from 0.78 to 0.83.
The total-only specification had maximum $\hat R=1.0006$, minimum bulk ESS $10347$, and minimum tail ESS $18462$. There were 0 divergences and 0 maximum-treedepth hits; chain E-BFMI ranged from 0.78 to 0.79.

The posterior medians (95\% credible intervals; Human alone, then Human+AI) were $b=0.32$ $[0.17,0.52]$ and $6.26$ $[4.85,7.94]$ for the report shift; $\sigma=2.10$ $[1.70,2.72]$ and $8.24$ $[6.62,10.17]$ for compression; $\tau=0.85$ $[0.66,1.15]$ and $3.82$ $[3.04,4.76]$ for shared person-offset variation; and $c=0.06$ $[-0.03,0.15]$ and $-0.62$ $[-1.06,-0.23]$ for the global-report shift. These quantities describe the latent report scale.

The reference slope $\varphi(b_k/\sigma_k)/[\sigma_k\varphi(0)]$ was $0.47$ $[0.36,0.58]$ in Human alone and $0.09$ $[0.07,0.11]$ in Human+AI. It fixes block difficulty and person/global offsets at zero, so it differs from the observed-score and IV slopes. Likewise, $c_k$ is not a pure aggregation effect: applying a nonlinear link after averaging block difficulties does not equal summing the four linked block expectations.

We therefore tested global estimates minus the sum of the four block estimates directly. The discrepancy was $-0.84$ points in Human+AI, CI $[-1.50,-0.18]$, $t(178)=-2.50$, $p=.013$, $d_z=-0.19$, and $+0.42$ in Human alone, CI $[-0.18,1.03]$, $t(186)=1.38$, $p=.170$, $d_z=0.10$. The group difference was $-1.26$, CI $[-2.15,-0.37]$, Welch's $t(359.59)=-2.78$, $p=.006$, $g=-0.29$.

\begin{figure}[!htp]
\centering
\includegraphics[width=0.9\linewidth]{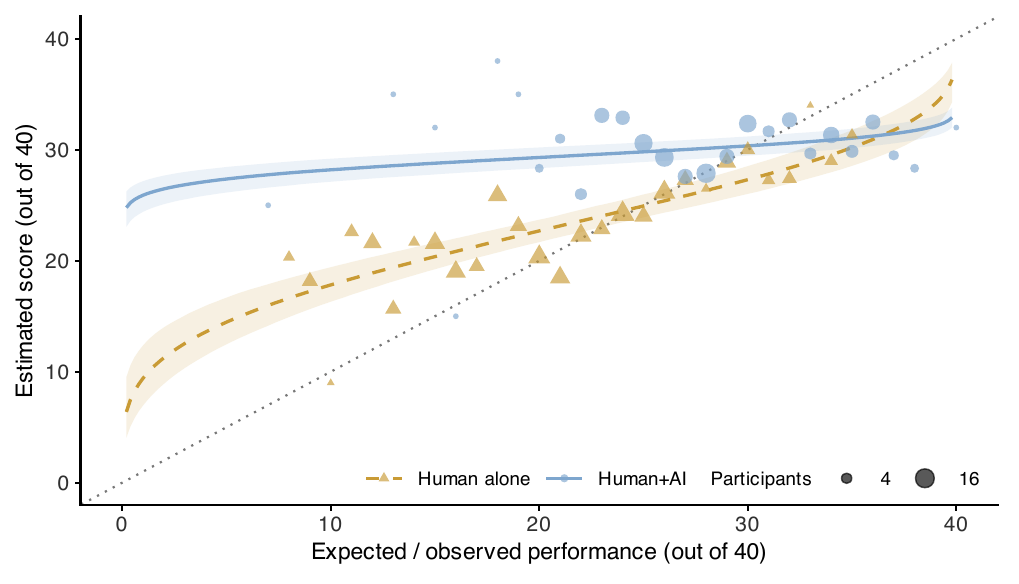}
\caption{Smooth group-level expected global estimates against expected achieved performance, integrating over the fitted person-offset distribution. Shaded bands indicate pointwise 95\% credible intervals for the mean curve, not prediction intervals for individual reports. Symbols show observed mean estimates at each achieved score, sized by participant count. The diagonal indicates agreement. Observed scores and expected performance are distinct quantities.}
\Description{Smooth expected global-estimate curves against expected achieved performance, with separate observed-score cell means overlaid. Bands show credible intervals for the mean curves and point sizes show participant counts.}
\label{fig:achperc}
\end{figure}

Figure~\ref{fig:achperc} shows smooth group-level curves obtained by varying latent skill and integrating over the fitted person-offset distribution. The horizontal coordinate is expected achieved performance (the sum of the four block expectations for the global curve), not observed score conditioned back to skill. Observed score-cell means are overlaid separately. Quartile checks below retain exactly the descriptive analysis's participant assignments.

\begin{figure}[!ht]
\centering
\includegraphics[width=0.9\linewidth]{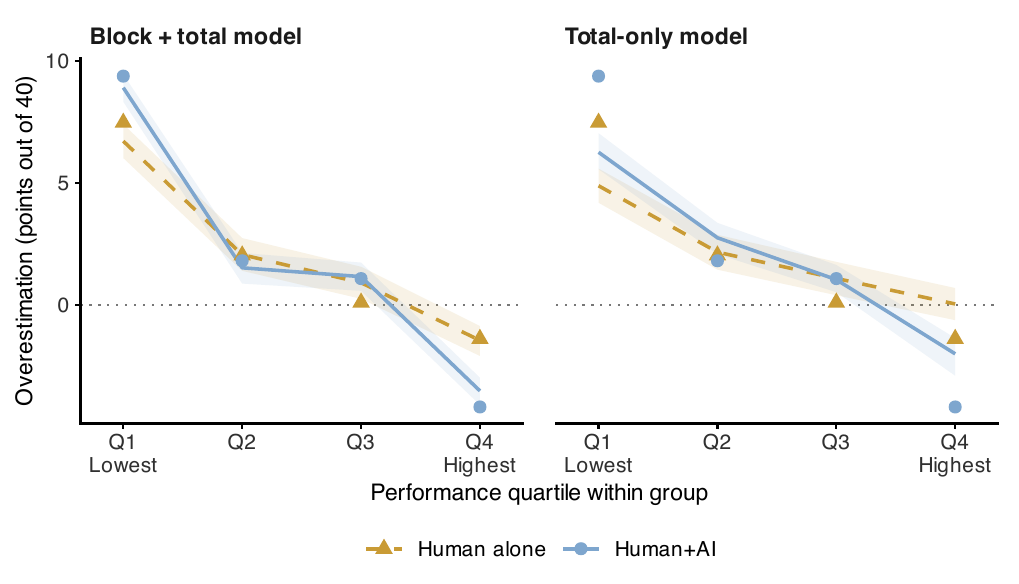}
\caption{Observed quartile overestimation (points) and conditional posterior expected overestimation (lines and 95\% credible bands) for the extended (left) and total-only (right) computational specifications. Quartile membership matches the descriptive analysis.}
\Description{Observed and fitted overestimation across within-group quartiles.}
\label{fig:crossing}
\end{figure}

Figure~\ref{fig:crossing} compares the observed quartile contrasts with conditional predictions from both specifications; the extended specification more closely follows the observed contrasts. Exact values are reported in Appendix~\ref{app:computational_checks}.
These are conditional, in-sample checks, not out-of-sample comparisons. Replicated-report checks identified distributional misfit: Human+AI mental-rotation estimates reached 10/10 for $33\%$ of participants, compared with a posterior predictive median of $17\%$ (95\% interval $[12,22]\%$); for matrix reasoning the corresponding values were $28\%$ versus $16\%$ $[11,21]\%$. The specification therefore respects the bounds but does not fully reproduce the ceiling mass or all block means and spreads\footnote{Separate fits with block-report offsets or beta-binomial reports yielded similar conditional quartile contrasts (posterior medians across specifications: 12.26--12.55 points for Human+AI and 8.16--8.39 for Human alone). Block offsets improved mean fit; beta-binomial reports improved some ceiling frequencies but overpredicted variability in several blocks. Neither eliminated the distributional misfit, and the low-performance portion of the Human+AI mean curve was specification-sensitive. We therefore retain the primary specification. }.

\subsubsection{Item-level accuracy and confidence}
\label{sec:res_mechanism}

We regressed item correctness (binomial-logit) and item confidence (linear) on standardized AI model performance, separately by group. Participant, block, and item were modeled as crossed random effects; the AI model's accuracy varies between items, not within them. 

\begin{figure}[t]
\centering
\includegraphics[width=0.9\linewidth]{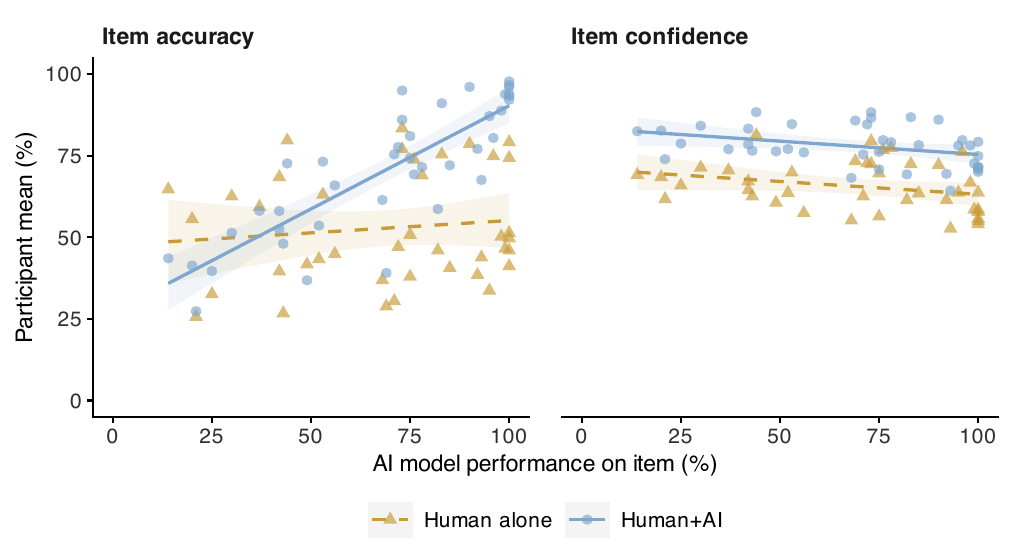}
  \caption{Participants' mean accuracy (left) and mean confidence (right), in percent, on each of the 40 items against the accuracy of the AI model alone on that item (out of 100 runs), for Human alone (yellow) and Human+AI (blue). Each point represents one item. Lines and 95\% confidence bands are unadjusted descriptive linear fits to item means, not the crossed mixed-model estimates reported in the text. Accuracy has a positive association with AI model accuracy in Human+AI; the confidence association is substantially shallower.
  }
  \Description{Item-level accuracy and confidence plotted against the AI model's item-level accuracy.}
  \label{fig:mechanism}
\end{figure}

Human+AI accuracy had a positive association with AI model performance. A change from the mean to one SD above the mean of AI model item performance corresponded to $5.06$ points out of 40 (approximate 95\% CI $[3.42,6.96]$), compared with $0.04$ $[-0.67,0.75]$ for confidence. Human alone estimates were $1.55$ $[-1.25,4.21]$ for accuracy and $-0.08$ $[-1.09,0.93]$ for confidence (\autoref{fig:mechanism}). Accuracy contrasts set random effects to zero; these are not population-marginal effects. The comparison of accuracy and confidence is descriptive, not a direct statistical test of their difference.

Consistent with this pattern, item-level confidence-minus-accuracy was negatively associated with AI model performance in Human+AI, $r=-0.89$ (95\% CI $[-0.94,-0.80]$), whereas the Human alone association was uncertain, $r=-0.29$ $[-0.55,0.03]$. Approximately $89\%$ of the Human+AI covariance arose algebraically from the negative accuracy term, so this is not independent evidence for a mechanism.

\clearpage

Pooled interaction sensitivities are reported in Appendix~\ref{app:item_sensitivity}. The confidence specification had a singular random-effects fit; separate accuracy and confidence interactions do not constitute a direct test of their difference. Uncertainty is conditional on the 100-run benchmark.

\subsubsection{Confidence and aggregation}
\label{sec:res_crossing}

The second part of RQ3 decomposed overestimation into item-level confidence error and the discrepancy between summed confidence and the whole-battery estimate.
Overestimation on the whole battery is calculated as the whole-battery estimate minus the achieved score, and it is divided into two parts that are summed. 

First, we divided each of the 40 confidence ratings, given on a slider from 0 to 100, by 100 so each rating could be interpreted as the probability that the answer is correct, and summed the 40 probabilities into the implied number of correct answers. 
Subtracting the achieved score from this implied total gives the confidence component, that is, the overestimation already contained in the item-level judgments. 
Second, subtracting the implied total from the whole-battery estimate yields the aggregation component, which is the amount by which the whole-battery estimate differs from what the item confidences imply. Note that the second component describes a relation between two reports and not a cognitive operation we observed participants perform.

To describe the performance association of each component, we regressed each on achieved score.
The confidence component decreased by $0.633$ points for each additional point of performance (95\% CI $[-0.780, -0.486]$) in the Human alone group and by $-0.850$ ($[-1.015, -0.686]$) in the Human+AI group.
The aggregation component showed no clear association with performance, with coefficients of $0.094$, $[-0.037, 0.224]$ (Human alone) and $-0.072$, $[-0.234, 0.090]$ (Human+AI). The two slopes sum to the overall overestimation slope, $-0.539$ in the Human alone group and $-0.923$ in the Human+AI group. Most of the observed negative association is located descriptively in the confidence-minus-score component. Subtracting achieved score contributes mechanically to that slope.

\FloatBarrier

\section{Discussion}

We examined how people judge reasoning performance when working alone or with an AI model across four task families. Human+AI participants achieved higher scores than Human alone, but their self-estimates tracked those scores weakly. Overall synergy beyond the AI model alone remained uncertain (RQ1). Similar average estimation errors concealed lower pooled confidence discrimination in Human+AI, with a smaller, uncertain within-block difference (RQ2), and a larger observed DKE quartile contrast (RQ3). The DKE pattern remained within each group under disjoint-score controls, although the controlled difference between groups was uncertain. The central finding is a disconnect between successful task performance and knowing how well the work was done.

\subsection{Assessing a partnership, not only oneself}

AI assistance affects both performance and the evidence self-assessment draws on. A Human+AI score reflects the human, the AI model, and how the two performed together on a particular task. In our study, the pair outperformed Human alone overall, and it outperformed the AI model alone on matrix reasoning, while overall synergy was not established. This distinction between augmentation and synergy aligns with \citet{fernandes_ai_2024} and \citet{vaccaro2024combinations}. Moreover, it modifies what a low achieved score represents. A lower Human+AI score does not imply that the human contributed less. Therefore, a participant who wants to know how well they performed must evaluate what they and this AI model accomplish jointly on this task, which is different from knowing their own or the AI model's abilities.

Knowing the two parts separately is not enough to assess the pair. Human+AI participants' estimates of the AI model were accurate on average, yet they weakly followed its performance across blocks (with a coefficient of only 0.324). Human alone participants expected the pair to exceed the higher of its two parts, which is in line with inflated expectations of AI benefits \cite{kloft2023ai, cave2019hopes, villa2023placebo, kosch2023placebo}. Their estimate of the pair followed their estimate of the AI model more closely than their estimate of themselves, and 24.6\% of their reports placed the pair above both parts. Among Human+AI participants, who estimated the pair they had actually worked as, the estimate of the pair followed both parts similarly. These associations motivate studying how people form expectations of collaboration with AI, beyond whether they know either part separately.

Participants also rated themselves against other participants on a scale of 0 to 100, which requires a rank rather than a count and thus does not exceed the maximum of 40. 
Human+AI participants rated their performance and reasoning ability with the AI model higher than Human-only participants, though both groups overestimated their total score by about two points.
A relative judgment therefore showed a group difference that the signed score error did not. However, this does not correct for the ceiling. Participants were asked to compare themselves to other participants in this study, which does not specify whether this means their own group or both groups, and Human+AI participants did indeed rank higher in the pooled sample. Absolute score error and relative placement should be compared to the population they refer to \cite{moore2008trouble}.

\subsection{Average bias conceals the distribution of self-assessment errors}

A small average error does not establish accurate individual self-assessment. Both groups' mean absolute errors were about six points, despite signed mean errors of about two. The exploratory equivalence checks supported small group differences in both measures at the post-hoc $|d|<.30$ margin. The larger average overestimation with AI reported in Study~1 of \citet{fernandes_ai_2024} was therefore not replicated here. 
The more significant distinction was how estimates varied with performance. Across observed-score quartiles, Human+AI estimates remained near 30 out of 40 points. In a group mean, underestimation by higher performers and overestimation by lower performers can cancel each other.

The DKE makes that visible. Both groups showed greater overestimation in the lowest than in the highest quartile; underestimation in the highest quartile was established only for Human+AI. The observed contrast was larger in Human+AI than in Human alone, while the controlled group difference was uncertain. These findings extend the unresolved evidence from \citet{fernandes_ai_2024}, whose AI users showed performance-dependent overestimation without meeting their DKE criterion. With more measurement under varying tasks, our results establish that a DKE pattern in Human+AI can persist beyond descriptive quartile comparisons.

Score-noise controls reduced but did not eliminate the within-group pattern. A 40-item score is an imperfect measure of ability, and ranking participants on the same score that is subtracted from their estimate produces some of the difference between quartiles on its own. The difference persisted when one half of the items was used to rank participants, and the remaining half entered their estimation error. A simulation with constant bias and constant noise produced smaller contrasts than the observed ones, and correcting for measurement error left Human+AI estimates only weakly related to their scores \cite{krueger2002unskilled, feld2017estimating, gignac2020dunning}. 

These checks support performance-dependent estimation error within each group under the specified controls. The computational extension combines block and global reports into a single description of this association. It more closely reproduces the observed conditional quartile contrasts than a total-only specification\footnote{The completed WAMBS comparisons found stable score-level predictions under longer sampling and alternative priors (Appendix~\ref{app:bayesian_diagnostics}).}.

\subsection{Monitoring across tasks and within answers}

Knowing how many answers are correct differs from knowing which answers are correct. Human+AI participants were more confident in each answer, but confidence separated correct from incorrect answers less effectively. The difference was substantially smaller when comparisons were restricted to the same block. About 80\% of the correct and incorrect pairs behind the pooled measure crossed blocks. Metacognition in human-AI interaction should therefore distinguish task-level assessments from answer-level discrimination \cite{tankelevitch2023metacognitive, fleming_how_2014, fleming2024metacognition}. 

The item analysis is consistent with this distinction. Human+AI accuracy was more closely related to benchmarked AI model performance than human accuracy alone, but the confidence interaction remained uncertain. In parallel, the overestimation decomposition placed most of the negative score association in summed item confidence minus achieved score. Global estimates differed from summed item confidence. Human+AI participants estimated their total below the sum of their block estimates.

A possible explanation for this disconnect is a change in the usefulness of self-assessment cues. In the cue-utilization account of metacognition, people infer how well they know something from available cues rather than directly observing their knowledge \cite{koriat1997monitoring}. With AI assistance, the ease of reaching an answer may reflect the availability of a suggested solution rather than successful reasoning or verification. Cues that are informative during independent problem solving may therefore become less diagnostic of joint performance. From this perspective, the DKE pattern could reflect a mismatch between the cues supporting self-assessment and the factors determining performance. This shifts the design question from simply reducing confidence to helping users recognize evidence that reliably indicates the quality of joint work. Note that this is \cite{fernandes2026explaining} who found that long reasoning traces can impair cognitive performance in HAI. Thus, which cues are utilized for metacogntive estimates in HAI need further research. 

This monitoring distinction is relevant for collaboration. According to theory, informative confidence from either agent can lead to synergy when combined optimally and with appropriate error dependence \cite{li2026metacognitive}. Our findings do not test the combination rule, but they do demonstrate why we should evaluate a confidence signal at the level at which it guides a decision. An overall confidence score may mask difficulty distinguishing tasks, whereas good average performance may mask uncertainty about specific answers. Effective monitoring requires evidence about both.

\subsection{Implications for HAI evaluation and design}

Evaluations should assess the quality of joint work and users' ability to evaluate it separately. This follows the performance--monitoring distinction raised by \citet{fernandes_ai_2024}, with our block-level results adding a task-specific dimension. Reporting average performance and signed bias alone would miss much of the present pattern. Distributional self-assessment errors, absolute error, and both pooled and within-task discrimination provide complementary evidence. The same-item benchmark and repeated block estimates help locate where beliefs and performance diverge \cite{klein2024performance}.

Interfaces should support recognizing when correction is needed. Lower-performing Human+AI participants may perceive less need for feedback than their scores warrant, so checks triggered only by users requesting help could miss useful opportunities. Conversely, blanket confidence reduction could worsen calibration among higher-performing participants who underestimate their scores. Task-specific feedback, verification prompts, and explain-back or cognitive forcing interventions are therefore candidates for evaluation \cite{forcing2021, fisher2021harder}. Their timing should be tested against both task risk and user confidence, with benefits weighed against effort and interruption.

Calibration also needs to be studied over time. The earlier work of \citet{fernandes_ai_2024} motivates examining feedback and AI literacy together, but its literacy associations do not establish that training is ineffective. Repeated interaction could help users learn the performance of a particular Human+AI partnership, or leave expectations unchanged despite errors. Experiments comparing literacy instruction, outcome feedback, and verification support could measure both immediate joint performance and later performance without assistance. This would distinguish learning to use an AI model from learning to judge the resulting work.

\subsection{Limitations}
\label{sec:limitations}

The study was exploratory, not preregistered, and recruited the two groups through separate calls rather than random assignment. Selection differences may contribute to group contrasts, so those contrasts do not identify causal effects of AI assistance. Four fixed task types also confound AI model performance with other task characteristics. They allow comparisons across this battery, not an isolated test of why self-assessment tracking differs or a broad inference to all reasoning tasks.

The DKE controls evaluate specified statistical accounts. Disjoint item halves need not have independent measurement errors, and split-half IV inference depends on its instrument assumptions. The simulation contrasts are not identified fractions of the observed DKE. Residual-spread tests concern between-person variation, not within-person metacognitive noise. These constraints prevent the remaining pattern from being attributed uniquely to a selective metacognitive deficit.

Several measures have important limits. Summing the Unsure--Certain ratings assumes a probability-of-correctness interpretation; subtracting achieved score introduces an algebraic association in the confidence decomposition. Global-minus-block estimates describe reporting differences, not an observed aggregation operation. AUROC depends on first-order performance \cite{rahnev2025comprehensive}, and perfect block scores leave it undefined. Human+AI ceilings were substantial, especially on syllogisms. The computational specification respects response bounds but underpredicts some ceiling proportions and other features of the report distribution. Longer sampling and stable prior-sensitivity findings do not repair that misfit or establish distinct psychological mechanisms.

Experienced and hypothetical judgments use different references. Human+AI participants' estimates of solo performance were compared with the Human alone group mean, not their own unobserved solo score. The reverse group's pair estimates were hypothetical. These comparisons cannot establish individual counterfactual error or rank how difficult the three targets are to judge. Component-belief regressions relate bounded, error-prone reports and do not recover cognitive weights or the direction of belief formation. Percentile judgments have a further ambiguity: the reference population was not explicitly restricted to the participant's group. Against pooled performance ranks, Human+AI showed slight underplacement and Human alone substantial overplacement, precluding a claim of greater ceiling-corrected bias in Human+AI.

The AI model benchmark used 100 runs on the same fixed items, collected after the participant study. It omitted instruction and practice context and reset history between blocks, unlike the live session. Its conditions were therefore not identical to participants' conditions, affecting the synergy comparison and analyses using benchmarked item performance. Original answer-extraction scores remain primary, with the confirmed correction reported as a sensitivity; the audit cannot rule out other extraction errors. Repeated runs improve precision on this battery but do not supply independent samples of tasks.

Finally, participants used one AI model in one session without accuracy feedback and had to prompt it at least once per item. We summarized prompt counts but did not code answer adoption, verification, or explanation evaluation. The accuracy--confidence pattern therefore does not identify how participants used the AI model or revised their beliefs. A within-person design with optional consultation, coded interaction strategies, and repeated feedback could examine those processes and separate judgments of current joint performance from independent ability.

\section{Conclusion}

Human+AI participants performed better than Human alone while judging that performance less consistently across score levels. Similar average bias concealed a larger observed DKE contrast and lower pooled confidence discrimination. These findings make self-assessment of joint work a distinct target for HAI design: systems should help users evaluate where a partnership succeeds and where its answers need checking, alongside helping them produce those answers.

\begin{acks}
This work was supported by the European Research Council (ERC) under the European Union's Horizon Europe research and innovation programme, AmplifAI (grant agreement No. 101217557). Views and opinions expressed are, however, those of the authors only and do not necessarily reflect those of the European Union or the European Research Council Executive Agency. Neither the European Union nor the granting authority can be held responsible for them.

This work is also funded by the Finnish Doctoral Program Network in Artificial Intelligence, AI-DOC (decision number VN/3137/2024-OKM-6).

\noindent\includegraphics[width=0.32\linewidth]{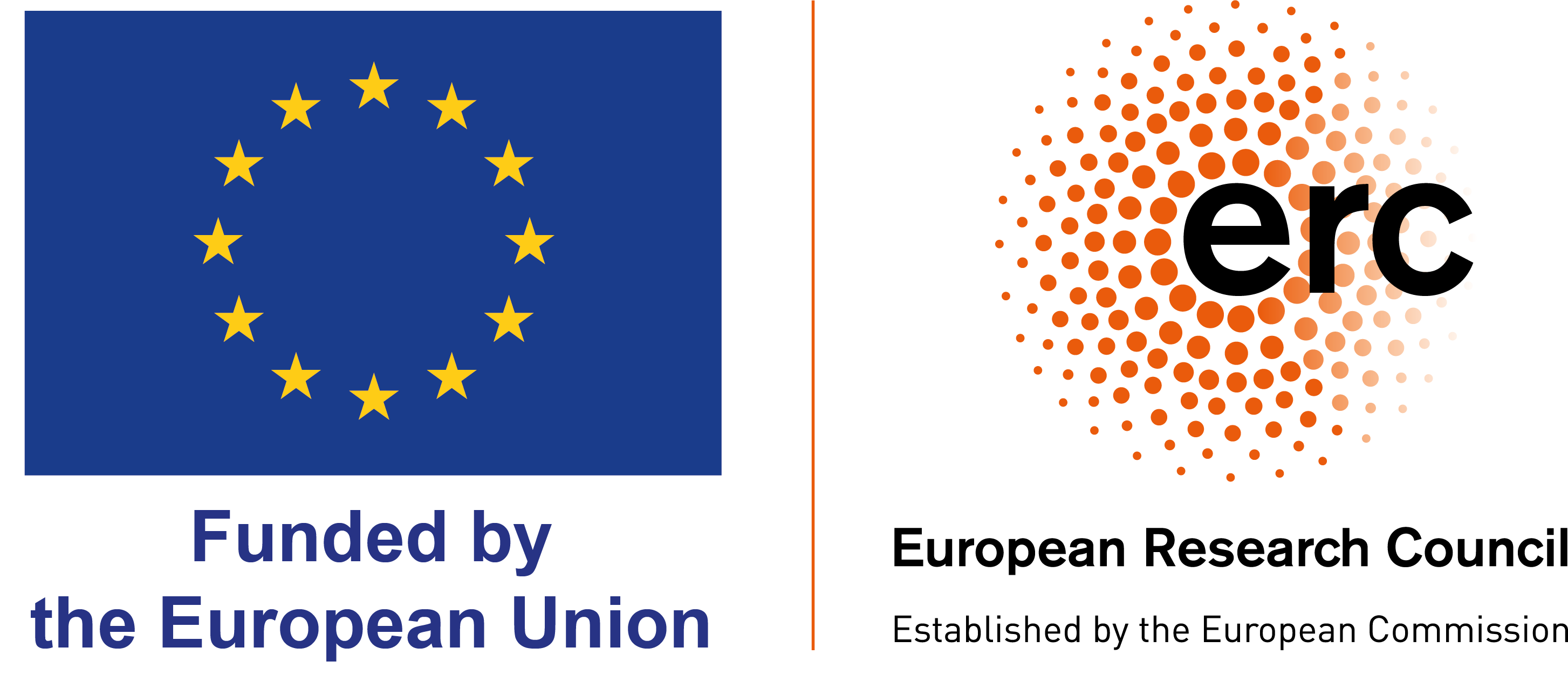}
\end{acks}
\bibliographystyle{ACM-Reference-Format}
\bibliography{preprint_references}

@article{Matzen2010,
  title = {Recreating Raven’s: Software for systematically generating large numbers of Raven-like matrix problems with normed properties},
  volume = {42},
  ISSN = {1554-3528},
  url = {http://dx.doi.org/10.3758/BRM.42.2.525},
  DOI = {10.3758/brm.42.2.525},
  number = {2},
  journal = {Behavior Research Methods},
  publisher = {Springer Science and Business Media LLC},
  author = {Matzen,  Laura E. and Benz,  Zachary O. and Dixon,  Kevin R. and Posey,  Jamie and Kroger,  James K. and Speed,  Ann E.},
  year = {2010},
  month = May,
  pages = {525–541}
}

@article{Shepard1971,
  title = {Mental Rotation of Three-Dimensional Objects},
  volume = {171},
  ISSN = {1095-9203},
  url = {http://dx.doi.org/10.1126/science.171.3972.701},
  DOI = {10.1126/science.171.3972.701},
  number = {3972},
  journal = {Science},
  publisher = {American Association for the Advancement of Science (AAAS)},
  author = {Shepard,  Roger N. and Metzler,  Jacqueline},
  year = {1971},
  month = Feb,
  pages = {701–703}
}

@misc{Lewis2024,
  doi = {10.48550/ARXIV.2411.14215},
  url = {https://arxiv.org/abs/2411.14215},
  author = {Lewis,  Martha and Mitchell,  Melanie},
  title = {Evaluating the Robustness of Analogical Reasoning in Large Language Models},
  publisher = {arXiv},
  year = {2024}
}

@misc{Webb2022,
  doi = {10.48550/ARXIV.2212.09196},
  url = {https://arxiv.org/abs/2212.09196},
  author = {Webb,  Taylor and Holyoak,  Keith J. and Lu,  Hongjing},
  title = {Emergent Analogical Reasoning in Large Language Models},
  publisher = {arXiv},
  year = {2022}
}

@article{carpenter_what_1990,
  title = {What one intelligence test measures: {A} theoretical account of the processing in the {Raven} {Progressive} {Matrices} {Test}.},
  volume = {97},
  issn = {1939-1471, 0033-295X},
  url = {https://doi.apa.org/doi/10.1037/0033-295X.97.3.404},
  doi = {10.1037/0033-295X.97.3.404},
  number = {3},
  journal = {Psychological Review},
  author = {Carpenter, Patricia A. and Just, Marcel A. and Shell, Peter},
  year = {1990},
  pages = {404--431}
}

@article{harris_measuring_2020,
  title = {Measuring {Intelligence} with the {Sandia} {Matrices}: {Psychometric} {Review} and {Recommendations} for {Free} {Raven}-{Like} {Item} {Sets}},
  volume = {6},
  issn = {23778822},
  url = {https://scholarworks.bgsu.edu/pad/vol6/iss3/6/},
  doi = {10.25035/pad.2020.03.006},
  number = {3},
  journal = {Personnel Assessment and Decisions},
  author = {Harris, Alexandra and McMillan, Jeremiah and Listyg, Benjamin and Matzen, Laura and Carter, Nathan},
  month = dec,
  year = {2020}
}

@article{khemlani_theories_2012,
  title = {Theories of the syllogism: {A} meta-analysis.},
  volume = {138},
  issn = {1939-1455, 0033-2909},
  url = {https://doi.apa.org/doi/10.1037/a0026841},
  doi = {10.1037/a0026841},
  number = {3},
  journal = {Psychological Bulletin},
  author = {Khemlani, Sangeet and Johnson-Laird, P. N.},
  year = {2012},
  pages = {427--457}
}

@inbook{hofstadter1995copycat,
  author = {Hofstadter, Douglas and Mitchell, Melanie},
  title = {The Copycat project: a model of mental fluidity and analogy-making},
  year = {1995},
  isbn = {0465051545},
  publisher = {Basic Books, Inc.},
  address = {USA},
  booktitle = {Fluid Concepts and Creative Analogies: Computer Models of the Fundamental Mechanisms of Thought},
  pages = {205–267},
  numpages = {63}
}

@InProceedings{pmlr-v80-barrett18a,
  title = {Measuring abstract reasoning in neural networks},
  author = {Barrett, David and Hill, Felix and Santoro, Adam and Morcos, Ari and Lillicrap, Timothy},
  booktitle = {Proceedings of the 35th International Conference on Machine Learning},
  pages = {511--520},
  year = {2018},
  editor = {Dy, Jennifer and Krause, Andreas},
  volume = {80},
  series = {Proceedings of Machine Learning Research},
  month = {10--15 Jul},
  publisher = {PMLR},
  url = {https://proceedings.mlr.press/v80/barrett18a.html},
  doi = {10.48550/arXiv.1807.04225}
}

@INPROCEEDINGS{Zhang_RAVEN,
  author = {Zhang, Chi and Gao, Feng and Jia, Baoxiong and Zhu, Yixin and Zhu, Song-Chun},
  booktitle = {2019 IEEE/CVF Conference on Computer Vision and Pattern Recognition (CVPR)},
  title = {RAVEN: A Dataset for Relational and Analogical Visual REasoNing},
  year = {2019},
  volume = {},
  number = {},
  pages = {5312-5322},
  doi = {10.1109/CVPR.2019.00546}
}

@misc{ramakrishnan_does_2024,
  title = {Does {Spatial} {Cognition} {Emerge} in {Frontier} {Models}?},
  url = {https://arxiv.org/abs/2410.06468},
  doi = {10.48550/ARXIV.2410.06468},
  publisher = {arXiv},
  author = {Ramakrishnan, Santhosh Kumar and Wijmans, Erik and Kraehenbuehl, Philipp and Koltun, Vladlen},
  year = {2024},
  note = {Version Number: 2}
}

@inproceedings{eisape-etal-2024-systematic,
  title = "A Systematic Comparison of Syllogistic Reasoning in Humans and Language Models",
  author = "Eisape, Tiwalayo  and
      Tessler, Michael  and
      Dasgupta, Ishita  and
      Sha, Fei  and
      van Steenkiste, Sjoerd  and
      Linzen, Tal",
  editor = "Duh, Kevin  and
      Gomez, Helena  and
      Bethard, Steven",
  booktitle = "Proceedings of the 2024 Conference of the North American Chapter of the Association for Computational Linguistics: Human Language Technologies (Volume 1: Long Papers)",
  month = jun,
  year = "2024",
  address = "Mexico City, Mexico",
  publisher = "Association for Computational Linguistics",
  url = "https://aclanthology.org/2024.naacl-long.466/",
  doi = "10.18653/v1/2024.naacl-long.466",
  pages = "8425--8444"
}

@inproceedings{ozeki-etal-2024-exploring,
  title = "Exploring Reasoning Biases in Large Language Models Through Syllogism: Insights from the {N}eu{BAROCO} Dataset",
  author = "Ozeki, Kentaro  and
      Ando, Risako  and
      Morishita, Takanobu  and
      Abe, Hirohiko  and
      Mineshima, Koji  and
      Okada, Mitsuhiro",
  editor = "Ku, Lun-Wei  and
      Martins, Andre  and
      Srikumar, Vivek",
  booktitle = "Findings of the Association for Computational Linguistics: ACL 2024",
  month = aug,
  year = "2024",
  address = "Bangkok, Thailand",
  publisher = "Association for Computational Linguistics",
  url = "https://aclanthology.org/2024.findings-acl.950/",
  doi = "10.18653/v1/2024.findings-acl.950",
  pages = "16063--16077"
}

@article{JohnsonLaird1984,
  title = {Syllogistic inference},
  volume = {16},
  ISSN = {0010-0277},
  url = {http://dx.doi.org/10.1016/0010-0277(84)90035-0},
  DOI = {10.1016/0010-0277(84)90035-0},
  number = {1},
  journal = {Cognition},
  publisher = {Elsevier BV},
  author = {Johnson-Laird,  P.N. and Bara,  Bruno G.},
  year = {1984},
  month = Feb,
  pages = {1–61}
}

@inproceedings{10.1145/3397481.3450639,
  author = {Nourani, Mahsan and Roy, Chiradeep and Block, Jeremy E and Honeycutt, Donald R and Rahman, Tahrima and Ragan, Eric and Gogate, Vibhav},
  title = {Anchoring Bias Affects Mental Model Formation and User Reliance in Explainable AI Systems},
  year = {2021},
  isbn = {9781450380171},
  publisher = {Association for Computing Machinery},
  address = {New York, NY, USA},
  url = {https://doi.org/10.1145/3397481.3450639},
  doi = {10.1145/3397481.3450639},
  booktitle = {Proceedings of the 26th International Conference on Intelligent User Interfaces},
  pages = {340–350},
  numpages = {11},
  location = {College Station, TX, USA},
  series = {IUI '21}
}

@inproceedings{10.1145/3411764.3445717,
  author = {Bansal, Gagan and Wu, Tongshuang and Zhou, Joyce and Fok, Raymond and Nushi, Besmira and Kamar, Ece and Ribeiro, Marco Tulio and Weld, Daniel},
  title = {Does the Whole Exceed its Parts? The Effect of AI Explanations on Complementary Team Performance},
  year = {2021},
  isbn = {9781450380966},
  publisher = {Association for Computing Machinery},
  address = {New York, NY, USA},
  url = {https://doi.org/10.1145/3411764.3445717},
  doi = {10.1145/3411764.3445717},
  booktitle = {Proceedings of the 2021 CHI Conference on Human Factors in Computing Systems},
  articleno = {81},
  numpages = {16},
  location = {Yokohama, Japan},
  series = {CHI '21}
}

@article{ackerman2017meta,
  title = {Meta-Reasoning: Monitoring and Control of Thinking and Reasoning},
  journal = {Trends in Cognitive Sciences},
  volume = {21},
  number = {8},
  pages = {607-617},
  year = {2017},
  issn = {1364-6613},
  doi = {10.1016/j.tics.2017.05.004},
  url = {https://www.sciencedirect.com/science/article/pii/S1364661317301055},
  author = {Rakefet Ackerman and Valerie A. Thompson}
}

@article{bastani2024generative,
  title = {Generative AI Can Harm Learning},
  author = {Bastani, Hamsa and Bastani, Osbert and Sungu, Alp and Ge, Haosen and Kabakc{\i}, {\"O}zge and Mariman, Rei},
  journal = {Available at SSRN 4895486},
  volume = { 4895486},
  year = 2024,
  doi = {10.2139/ssrn.4895486}
}

@inproceedings{bertrand2022cognitive,
  author = {Bertrand, Astrid and Belloum, Rafik and Eagan, James R. and Maxwell, Winston},
  title = {How Cognitive Biases Affect XAI-assisted Decision-making: A Systematic Review},
  year = {2022},
  isbn = {9781450392471},
  publisher = {Association for Computing Machinery},
  address = {New York, NY, USA},
  url = {https://doi.org/10.1145/3514094.3534164},
  doi = {10.1145/3514094.3534164},
  booktitle = {Proceedings of the 2022 AAAI/ACM Conference on AI, Ethics, and Society},
  pages = {78–91},
  numpages = {14},
  location = {Oxford, United Kingdom},
  series = {AIES '22}
}

@article{brown1986evaluations,
  author = {Brown, Jonathon D.},
  title = {Evaluations of self and others: Self-enhancement biases in social judgments.},
  journal = {Social Cognition},
  year = {1986},
  publisher = {Guilford Publications},
  address = {US},
  volume = {4},
  number = {4},
  pages = {353-376},
  doi = {10.1521/soco.1986.4.4.353},
  url = {https://doi.org/10.1521/soco.1986.4.4.353}
}

@article{burkner2017brms,
  title = {brms: An R Package for Bayesian Multilevel Models Using Stan},
  author = {B{\"u}rkner, Paul-Christian},
  year = 2017,
  month = aug,
  journal = {Journal of Statistical Software},
  volume = 80,
  number = 1,
  numpages = 28,
  doi = {10.18637/jss.v080.i01}
}

@article{burson2006skilled,
  title = {Skilled or unskilled, but still unaware of it: how perceptions of difficulty drive miscalibration in relative comparisons.},
  author = {Burson, Katherine A and Larrick, Richard P and Klayman, Joshua},
  journal = {Journal of personality and social psychology},
  volume = 90,
  number = 1,
  pages = 60,
  year = 2006,
  publisher = {American Psychological Association},
  doi = {10.1037/0022-3514.90.1.60}
}

@article{cave2019hopes,
  title = {Hopes and fears for intelligent machines in fiction and reality},
  author = {Cave, Stephen and Dihal, Kanta},
  journal = {Nature machine intelligence},
  volume = 1,
  number = 2,
  pages = {74--78},
  year = 2019,
  publisher = {Nature Publishing Group UK London},
  doi = {10.1038/s42256-019-0020-9}
}

@article{colombatto2025metacognition,
  title = {Metacognition and Confidence Dynamics in Advice Taking from Generative AI},
  author = {Colombatto, Clara and Rintel, Sean and Tankelevitch, Lev},
  journal = {arXiv preprint arXiv:2510.26508},
  year = {2025},
  doi = {10.48550/arXiv.2510.26508}
}

@article{draxlerghost24,
  author = {Draxler, Fiona and Werner, Anna and Lehmann, Florian and Hoppe, Matthias and Schmidt, Albrecht and Buschek, Daniel and Welsch, Robin},
  title = {The AI Ghostwriter Effect: When Users do not Perceive Ownership of AI-Generated Text but Self-Declare as Authors},
  year = 2024,
  publisher = {Association for Computing Machinery},
  address = {New York, NY, USA},
  volume = 31,
  number = 2,
  issn = {1073-0516},
  url = {https://doi.org/10.1145/3637875},
  doi = {10.1145/3637875},
  journal = {ACM Trans. Comput.-Hum. Interact.},
  month = {2},
  articleno = 25,
  numpages = 40
}

@incollection{dunning2011dunning,
  title = {The Dunning--Kruger effect: On being ignorant of one's own ignorance},
  author = {Dunning, David},
  booktitle = {Advances in experimental social psychology},
  volume = 44,
  doi = {10.1016/B978-0-12-385522-0.00005-6},
  pages = {247--296},
  year = 2011,
  publisher = {Elsevier}
}

@article{ehrlinger2008unskilled,
  title = {Why the unskilled are unaware: Further explorations of (absent) self-insight among the incompetent},
  journal = {Organizational Behavior and Human Decision Processes},
  volume = {105},
  number = {1},
  pages = {98-121},
  year = {2008},
  issn = {0749-5978},
  doi = {https://doi.org/10.1016/j.obhdp.2007.05.002},
  url = {https://www.sciencedirect.com/science/article/pii/S074959780700060X},
  author = {Joyce Ehrlinger and Kerri Johnson and Matthew Banner and David Dunning and Justin Kruger}
}

@inproceedings{eiband2019impact,
  title = {The impact of placebic explanations on trust in intelligent systems},
  author = {Eiband, Malin and Buschek, Daniel and Kremer, Alexander and Hussmann, Heinrich},
  booktitle = {Extended abstracts of the 2019 CHI conference on human factors in computing systems},
  doi = {10.1145/3290607.3312787},
  pages = {1--6},
  year = 2019
}

@article{engelbart1962augmenting,
  title = {Augmenting human intellect: A conceptual framework},
  author = {Engelbart, Douglas C},
  year = 1962,
  journal = {Menlo Park, CA},
  pages = 21
}

@article{fernandes_ai_2024,
  title = {AI makes you smarter but none the wiser: The disconnect between performance and metacognition},
  journal = {Computers in Human Behavior},
  volume = {175},
  pages = {108779},
  year = {2026},
  issn = {0747-5632},
  doi = {10.1016/j.chb.2025.108779},
  url = {https://www.sciencedirect.com/science/article/pii/S0747563225002262},
  author = {Daniela Fernandes and Steeven Villa and Salla Nicholls and Otso Haavisto and Daniel Buschek and Albrecht Schmidt and Thomas Kosch and Chenxinran Shen and Robin Welsch}
}

@incollection{fiedler2019metacognition,
  title = {Metacognition: Monitoring and Controlling One’s Own Knowledge, Reasoning and Decisions},
  url = {https://heiup.uni-heidelberg.de/catalog/book/470/chapter/6669},
  doi = {10.17885/heiup.470.c6669},
  booktitle = {The Psychology of Human Thought: An Introduction},
  publisher = {Heidelberg University Publishing},
  author = {Fiedler, Klaus and Ackerman, Rakefet and Scarampi, Chiara},
  editor = {Sternberg, Robert J. and Funke, Joachim},
  year = {2019},
  month = {7},
  pages = {89–111}
}

@article{fisher2021harder,
  title = {Harder than you think: How outside assistance leads to overconfidence},
  author = {Fisher, Matthew and Oppenheimer, Daniel M},
  journal = {Psychological Science},
  volume = {32},
  number = {4},
  pages = {598--610},
  year = {2021},
  publisher = {Sage Publications Sage CA: Los Angeles, CA},
  doi = {10.1177/0956797620975779}
}

@article{fleming2024metacognition,
  title = {Metacognition and confidence: A review and synthesis},
  author = {Fleming, Stephen},
  journal = {Annual Review of Psychology},
  volume = 75,
  doi = {10.1146/annurev-psych-022423-032425},
  pages = {241--268},
  year = 2024,
  publisher = {Annual Reviews}
}

@article{fleming_how_2014,
  title = {How to measure metacognition},
  volume = {8},
  doi = {10.3389/fnhum.2014.00443},
  journal = {Frontiers in Human Neuroscience},
  author = {Fleming, Stephen and Lau, Hakwan},
  month = jul,
  year = {2014},
  pages = {443}
}

@article{forcing2021,
  author = {Bu\c{c}inca, Zana and Malaya, Maja Barbara and Gajos, Krzysztof Z.},
  title = {To Trust or to Think: Cognitive Forcing Functions Can Reduce Overreliance on AI in AI-assisted Decision-making},
  year = 2021,
  publisher = {Association for Computing Machinery},
  address = {New York, NY, USA},
  volume = 5,
  number = {Cscw1},
  url = {https://doi.org/10.1145/3449287},
  doi = {10.1145/3449287},
  journal = {Proc. ACM Hum.-Comput. Interact.},
  month = {4},
  articleno = 188,
  numpages = 21
}

@article{gignac2020dunning,
  title = {The Dunning-Kruger effect is (mostly) a statistical artefact: Valid approaches to testing the hypothesis with individual differences data},
  author = {Gignac, Gilles E and Zajenkowski, Marcin},
  journal = {Intelligence},
  volume = 80,
  pages = 101449,
  doi = {10.1016/j.intell.2020.101449},
  year = 2020,
  publisher = {Elsevier}
}

@inproceedings{he2023knowing,
  author = {He, Gaole and Kuiper, Lucie and Gadiraju, Ujwal},
  title = {Knowing About Knowing: An Illusion of Human Competence Can Hinder Appropriate Reliance on AI Systems},
  year = {2023},
  isbn = {9781450394215},
  publisher = {Association for Computing Machinery},
  address = {New York, NY, USA},
  url = {https://doi.org/10.1145/3544548.3581025},
  doi = {10.1145/3544548.3581025},
  booktitle = {Proceedings of the 2023 CHI Conference on Human Factors in Computing Systems},
  articleno = {113},
  numpages = {18},
  location = {Hamburg, Germany},
  series = {CHI '23}
}

@article{jansen2021rational,
  title = {A rational model of the Dunning--Kruger effect supports insensitivity to evidence in low performers},
  author = {Jansen, Rachel A and Rafferty, Anna N and Griffiths, Thomas L},
  journal = {Nature Human Behaviour},
  volume = 5,
  number = 6,
  pages = {756--763},
  year = 2021,
  publisher = {Nature Publishing Group UK London},
  doi = {10.1038/s41562-021-01057-0}
}

@article{magnus2022statistical,
  author = {Magnus, Jan R. and Peresetsky, Anatoly A.},
  title = {A Statistical Explanation of the {D}unning--{K}ruger Effect},
  journal = {Frontiers in Psychology},
  volume = {13},
  pages = {840180},
  year = {2022},
  doi = {10.3389/fpsyg.2022.840180}
}

@article{johnson_source_1993,
  title = {Source monitoring.},
  volume = {114},
  issn = {1939-1455, 0033-2909},
  url = {https://doi.apa.org/doi/10.1037/0033-2909.114.1.3},
  doi = {10.1037/0033-2909.114.1.3},
  number = {1},
  journal = {Psychological Bulletin},
  author = {Johnson, Marcia K. and Hashtroudi, Shahin and Lindsay, D. Stephen},
  year = {1993},
  note = {Publisher: American Psychological Association (APA)},
  pages = {3--28}
}

@article{kelemen2000individual,
  title = {Individual differences in metacognition: Evidence against a general metacognitive ability},
  author = {Kelemen, William L and Frost, Peter J and Weaver, Charles A},
  journal = {Memory \& cognition},
  volume = 28,
  pages = {92--107},
  year = 2000,
  publisher = {Springer},
  doi = {https://doi.org/10.3758/BF03211579}
}

@inproceedings{kim2024m,
  author = {Kim, Sunnie S. Y. and Liao, Q. Vera and Vorvoreanu, Mihaela and Ballard, Stephanie and Vaughan, Jennifer Wortman},
  title = {"I'm Not Sure, But...": Examining the Impact of Large Language Models' Uncertainty Expression on User Reliance and Trust},
  year = {2024},
  isbn = {9798400704505},
  publisher = {Association for Computing Machinery},
  address = {New York, NY, USA},
  url = {https://doi.org/10.1145/3630106.3658941},
  doi = {10.1145/3630106.3658941},
  booktitle = {Proceedings of the 2024 ACM Conference on Fairness, Accountability, and Transparency},
  pages = {822–835},
  numpages = {14},
  location = {Rio de Janeiro, Brazil},
  series = {FAccT '24}
}

@article{klein2024performance,
  title = {The performance of ChatGPT and Bing on a computerized adaptive test of verbal intelligence},
  author = {Klein, Bal{\'a}zs and Kovacs, Kristof},
  journal = {PloS one},
  volume = 19,
  number = 7,
  pages = {e0307097},
  year = 2024,
  publisher = {Public Library of Science San Francisco, CA USA},
  doi = { 10.1371/journal.pone.0307097}
}

@article{klingbeil2024trust,
  title = {Trust and reliance on AI—An experimental study on the extent and costs of overreliance on AI},
  author = {Klingbeil, Artur and Gr{\"u}tzner, Cassandra and Schreck, Philipp},
  journal = {Computers in Human Behavior},
  volume = 160,
  pages = 108352,
  year = 2024,
  publisher = {Elsevier},
  doi = {10.1016/j.chb.2024.108352}
}

@inproceedings{kloft2023ai,
  author = {Kloft, Agnes Mercedes and Welsch, Robin and Kosch, Thomas and Villa, Steeven},
  title = {"AI enhances our performance, I have no doubt this one will do the same": The Placebo effect is robust to negative descriptions of AI},
  year = 2024,
  isbn = 9798400703300,
  publisher = {Association for Computing Machinery},
  address = {New York, NY, USA},
  url = {https://doi.org/10.1145/3613904.3642633},
  doi = {10.1145/3613904.3642633},
  booktitle = {Proceedings of the CHI Conference on Human Factors in Computing Systems},
  articleno = 299,
  numpages = 24,
  location = {Honolulu, HI, USA},
  series = {Chi '24}
}

@inproceedings{Kobiella2024,
  author = {Kobiella, Charlotte and Flores L\'{o}pez, Yarhy Said and Waltenberger, Franz and Draxler, Fiona and Schmidt, Albrecht},
  title = {"If the Machine Is As Good As Me, Then What Use Am I?" – How the Use of ChatGPT Changes Young Professionals' Perception of Productivity and Accomplishment},
  year = {2024},
  isbn = {9798400703300},
  publisher = {Association for Computing Machinery},
  address = {New York, NY, USA},
  url = {https://doi.org/10.1145/3613904.3641964},
  doi = {10.1145/3613904.3641964},
  booktitle = {Proceedings of the CHI Conference on Human Factors in Computing Systems},
  articleno = {1018},
  numpages = {16},
  location = {Honolulu, HI, USA},
  series = {CHI '24}
}

@article{koriat1997monitoring,
  title = {Monitoring one's own knowledge during study: A cue-utilization approach to judgments of learning.},
  author = {Koriat, Asher},
  journal = {Journal of experimental psychology: General},
  volume = {126},
  number = {4},
  pages = {349},
  year = {1997},
  publisher = {American Psychological Association},
  doi = {https://doi.org/10.1037/0096-3445.126.4.349}
}

@article{kosch2023placebo,
  title = {The Placebo Effect of Artificial Intelligence in Human--Computer Interaction},
  author = {Kosch, Thomas and Welsch, Robin and Chuang, Lewis and Schmidt, Albrecht},
  journal = {ACM Transactions on Computer-Human Interaction},
  volume = 29,
  doi = {10.1145/3529225},
  number = 6,
  pages = {1--32},
  year = 2023,
  publisher = {ACM New York, NY}
}

@article{kruger1999unskilled,
  title = {Unskilled and unaware of it: how difficulties in recognizing one's own incompetence lead to inflated self-assessments.},
  author = {Kruger, Justin and Dunning, David},
  journal = {Journal of personality and social psychology},
  volume = 77,
  number = 6,
  pages = 1121,
  year = 1999,
  publisher = {American Psychological Association},
  doi = {10.1037/0022-3514.77.6.1121}
}

@inproceedings{lu2021human,
  title = {Human reliance on machine learning models when performance feedback is limited: Heuristics and risks},
  author = {Lu, Zhuoran and Yin, Ming},
  booktitle = {Proceedings of the 2021 CHI Conference on Human Factors in Computing Systems},
  pages = {1--16},
  doi = {10.1145/3411764.3445562},
  year = 2021
}

@inproceedings{ma2023who,
  author = {Ma, Shuai and Lei, Ying and Wang, Xinru and Zheng, Chengbo and Shi, Chuhan and Yin, Ming and Ma, Xiaojuan},
  title = {Who Should I Trust: AI or Myself? Leveraging Human and AI Correctness Likelihood to Promote Appropriate Trust in AI-Assisted Decision-Making},
  year = {2023},
  isbn = {9781450394215},
  publisher = {Association for Computing Machinery},
  address = {New York, NY, USA},
  url = {https://doi.org/10.1145/3544548.3581058},
  doi = {10.1145/3544548.3581058},
  booktitle = {Proceedings of the 2023 CHI Conference on Human Factors in Computing Systems},
  articleno = {759},
  numpages = {19},
  location = {Hamburg, Germany},
  series = {CHI '23}
}

@article{okamura2020empirical,
  author = {Okamura, Kazuo and Yamada, Seiji},
  journal = {IEEE Access},
  title = {Empirical Evaluations of Framework for Adaptive Trust Calibration in Human-AI Cooperation},
  year = {2020},
  volume = {8},
  number = {},
  pages = {220335-220351},
  doi = {10.1109/ACCESS.2020.3042556}
}

@article{rafner2022deskilling,
  title = {Deskilling, upskilling, and reskilling: a case for hybrid intelligence},
  author = {Rafner, Janet and Dellermann, Dominik and Hjorth, Arthur and Veraszto, Dora and Kampf, Constance and MacKay, Wendy and Sherson, Jacob},
  journal = {Morals \& Machines},
  volume = 1,
  number = 2,
  pages = {24--39},
  doi = {10.5771/2747-5174-2021-2-24},
  year = 2022,
  publisher = {Nomos Verlagsgesellschaft mbH \& Co. KG}
}

@article{rahnev2025comprehensive,
  title = {A comprehensive assessment of current methods for measuring metacognition},
  author = {Rahnev, Dobromir},
  journal = {Nature Communications},
  volume = {16},
  number = {1},
  pages = {701},
  year = {2025},
  publisher = {Nature Publishing Group UK London},
  doi = {10.1038/s41467-025-56117-0}
}

@inproceedings{ramesh2026metacognitive,
  author = {Ramesh, Shri Harini and Daneshzand, Foroozan and Rashidi, Babak and Raj, Shriti and Subramonyam, Hariharan and Rajabiyazdi, Fateme},
  title = {Metacognitive Demands and Strategies While Using Off-The-Shelf AI Conversational Agents for Health Information Seeking},
  year = {2026},
  isbn = {9798400722783},
  publisher = {Association for Computing Machinery},
  address = {New York, NY, USA},
  url = {https://doi.org/10.1145/3772318.3791647},
  doi = {10.1145/3772318.3791647},
  booktitle = {Proceedings of the 2026 CHI Conference on Human Factors in Computing Systems},
  articleno = {28},
  numpages = {16},
  series = {CHI '26}
}

@article{rozenblit2002misunderstood,
  title = {The misunderstood limits of folk science: an illusion of explanatory depth},
  volume = {26},
  issn = {0364-0213},
  url = {https://www.sciencedirect.com/science/article/pii/S0364021302000782},
  number = {5},
  journal = {Cognitive Science},
  author = {Rozenblit, Leonid and Keil, Frank},
  year = {2002},
  pages = {521--562}
}

@book{shekar2024people,
  title = {People over trust AI-generated medical responses and view them to be as valid as doctors, despite low accuracy},
  author = {Shekar, Shruthi and Pataranutaporn, Pat and Sarabu, Chethan and Cecchi, Guillermo A and Maes, Pattie},
  journal = {arXiv preprint arXiv:2408.15266},
  doi = { 10.48550/arXiv.2408.15266},
  year = 2024
}

@inproceedings{si2024large,
  title = "Large Language Models Help Humans Verify Truthfulness {--} Except When They Are Convincingly Wrong",
  author = "Si, Chenglei  and
      Goyal, Navita  and
      Wu, Tongshuang  and
      Zhao, Chen  and
      Feng, Shi  and
      Daum{\'e} Iii, Hal  and
      Boyd-Graber, Jordan",
  editor = "Duh, Kevin  and
      Gomez, Helena  and
      Bethard, Steven",
  booktitle = "Proceedings of the 2024 Conference of the North American Chapter of the Association for Computational Linguistics: Human Language Technologies (Volume 1: Long Papers)",
  month = jun,
  year = "2024",
  address = "Mexico City, Mexico",
  publisher = "Association for Computational Linguistics",
  url = "https://aclanthology.org/2024.naacl-long.81/",
  doi = "10.18653/v1/2024.naacl-long.81",
  pages = "1459--1474"
}

@article{stadler_cognitive_2024,
  title = {Cognitive ease at a cost: {LLMs} reduce mental effort but compromise depth in student scientific inquiry},
  volume = {160},
  issn = {0747-5632},
  url = {https://www.sciencedirect.com/science/article/pii/S0747563224002541},
  doi = {10.1016/j.chb.2024.108386},
  journal = {Computers in Human Behavior},
  author = {Stadler, Matthias and Bannert, Maria and Sailer, Michael},
  year = {2024},
  pages = {108386}
}

@article{steyvers_bayesian_2022,
  title = {Bayesian modeling of human–{AI} complementarity},
  volume = 119,
  url = {https://www.pnas.org/doi/abs/10.1073/pnas.2111547119},
  doi = {10.1073/pnas.2111547119},
  number = 11,
  journal = {Proceedings of the National Academy of Sciences},
  author = {Steyvers, Mark and Tejeda, Heliodoro and Kerrigan, Gavin and Smyth, Padhraic},
  month = mar,
  year = 2022,
  note = {Publisher: Proceedings of the National Academy of Sciences},
  pages = {e2111547119}
}

@inproceedings{tankelevitch2023metacognitive,
  title = {The Metacognitive Demands and Opportunities of Generative AI},
  volume = {57},
  url = {http://dx.doi.org/10.1145/3613904.3642902},
  DOI = {10.1145/3613904.3642902},
  booktitle = {Proceedings of the CHI Conference on Human Factors in Computing Systems},
  publisher = {ACM},
  author = {Tankelevitch, Lev and Kewenig, Viktor and Simkute, Auste and Scott, Ava Elizabeth and Sarkar, Advait and Sellen, Abigail and Rintel, Sean},
  year = {2024},
  month = may,
  pages = {1–24}
}

@article{tankelevitch2025understanding,
  title = {Understanding, Protecting, and Augmenting Human Cognition with Generative AI: A Synthesis of the CHI 2025 Tools for Thought Workshop},
  author = {Tankelevitch, Lev and Glassman, Elena L and He, Jessica and Kittur, Aniket and Lee, Mina and Palani, Srishti and Sarkar, Advait and Ramos, Gonzalo and Rogers, Yvonne and Subramonyam, Hari},
  journal = {arXiv preprint arXiv:2508.21036},
  year = {2025},
  doi = {10.48550/arXiv.2508.21036}
}

@article{vaccaro2024combinations,
  title = {When combinations of humans and AI are useful: A systematic review and meta-analysis},
  author = {Vaccaro, Michelle and Almaatouq, Abdullah and Malone, Thomas},
  journal = {Nature Human Behaviour},
  volume = {8},
  number = {12},
  pages = {2293--2303},
  year = {2024},
  publisher = {Nature Publishing Group},
  doi = {10.1038/s41562-024-02024-1}
}

@article{vasconcelos2023explanations,
  title = {Explanations can reduce overreliance on ai systems during decision-making},
  author = {Vasconcelos, Helena and J{\"o}rke, Matthew and Grunde-McLaughlin, Madeleine and Gerstenberg, Tobias and Bernstein, Michael S and Krishna, Ranjay},
  journal = {Proceedings of the ACM on Human-Computer Interaction},
  volume = 7,
  number = {Cscw1},
  doi = {10.1145/3579605},
  pages = {1--38},
  year = 2023,
  publisher = {ACM New York, NY, USA}
}

@article{villa2023placebo,
  title = {The placebo effect of human augmentation: Anticipating cognitive augmentation increases risk-taking behavior},
  author = {Villa, Steeven and Kosch, Thomas and Grelka, Felix and Schmidt, Albrecht and Welsch, Robin},
  journal = {Computers in Human Behavior},
  volume = 146,
  doi = {10.1016/j.chb.2023.107787},
  pages = 107787,
  year = 2023,
  publisher = {Elsevier}
}

@article{von2025knowing,
  title = {Knowing ({Not}) to {Know}: {Explainable} {Artificial} {Intelligence} and {Human} {Metacognition}},
  issn = {1047-7047, 1526-5536},
  doi = {10.1287/isre.2024.1431},
  journal = {Information Systems Research},
  author = {Von Zahn, Moritz and Liebich, Lena and Jussupow, Ekaterina and Hinz, Oliver and Bauer, Kevin},
  year = {2025},
  pages = {isre.2024.1431}
}

@inproceedings{wang2021explanations,
  title = {Are Explanations Helpful? A Comparative Study of the Effects of Explanations in AI-Assisted Decision-Making},
  author = {Wang, Xinru and Yin, Ming},
  year = {2021},
  month = {04},
  pages = {318-328},
  doi = {10.1145/3397481.3450650}
}

@article{wickens2015complacency,
  title = {Complacency and automation bias in the use of imperfect automation},
  author = {Wickens, Christopher D and Clegg, Benjamin A and Vieane, Alex Z and Sebok, Angelia L},
  journal = {Human factors},
  volume = 57,
  number = 5,
  pages = {728--739},
  year = 2015,
  publisher = {Sage Publications Sage CA: Los Angeles, CA}
}

@article{yang2024competence,
  title = {When competence and confidence are at odds: a cross-country examination of the Dunning--Kruger effect},
  author = {Yang Hansen, Kajsa and Thorsen, Cecilia and Radi{\v{s}}i{\'c}, Jelena and Peixoto, Francisco and Laine, Anu and Liu, Xin},
  journal = {European Journal of Psychology of Education},
  pages = {1--23},
  year = {2024},
  publisher = {Springer}
}

@article{zell2020better,
  title = {The better-than-average effect in comparative self-evaluation: A comprehensive review and meta-analysis.},
  author = {Zell, Ethan and Strickhouser, Jason E and Sedikides, Constantine and Alicke, Mark D},
  journal = {Psychological bulletin},
  volume = 146,
  number = 2,
  pages = 118,
  year = 2020,
  doi = {10.1037/bul0000218},
  publisher = {American Psychological Association}
}

@inproceedings{Zindulka2026AIMemoryGap,
  author = {Zindulka, Tim and Goller, Sven and Fernandes, Daniela and Welsch, Robin and Buschek, Daniel},
  title = {The {AI} Memory Gap: Users Misremember What They Created With {AI} or Without},
  booktitle = {Proceedings of the 2026 CHI Conference on Human Factors in Computing Systems},
  series = {CHI '26},
  year = {2026},
  location = {Barcelona, Spain},
  publisher = {Association for Computing Machinery},
  address = {New York, NY, USA},
  doi = {10.1145/3772318.3791494},
  isbn = {979-8-4007-2278-3}
}

@article{logg2019algorithm,
  author = {Logg, Jennifer M. and Minson, Julia A. and Moore, Don A.},
  title = {Algorithm appreciation: People prefer algorithmic to human judgment},
  journal = {Organizational Behavior and Human Decision Processes},
  volume = {151},
  pages = {90--103},
  year = {2019},
  doi = {10.1016/j.obhdp.2018.12.005}
}

@article{dietvorst2015algorithm,
  author = {Dietvorst, Berkeley J. and Simmons, Joseph P. and Massey, Cade},
  title = {Algorithm aversion: People erroneously avoid algorithms after seeing them err.},
  journal = {Journal of Experimental Psychology: General},
  volume = {144},
  number = {1},
  pages = {114--126},
  year = {2015},
  doi = {10.1037/xge0000033}
}

@article{krueger2002unskilled,
  title = {Unskilled, Unaware, or Both? The Better-than-Average Heuristic and Statistical Regression Predict Errors in Estimates of Own Performance},
  author = {Krueger, Joachim and Mueller, Ross A.},
  journal = {Journal of Personality and Social Psychology},
  volume = {82},
  number = {2},
  pages = {180--188},
  year = {2002},
  doi = {https://doi.org/10.1037/0022-3514.82.2.180}
}

@article{nuhfer2016random,
  title = {Random Number Simulations Reveal How Random Noise Affects the Measurements and Graphical Portrayals of Self-Assessed Competency},
  author = {Nuhfer, Edward and Cogan, Christopher and Fleisher, Steven and Gaze, Eric and Wirth, Karl},
  journal = {Numeracy},
  volume = {9},
  number = {1},
  pages = {4},
  year = {2016},
  doi = {http://dx.doi.org/10.5038/1936-4660.9.1.4}
}

@article{feld2017estimating,
  title = {Estimating the Relationship between Skill and Overconfidence},
  author = {Feld, Jan and Sauermann, Jan and de Grip, Andries},
  journal = {Journal of Behavioral and Experimental Economics},
  volume = {68},
  pages = {18--24},
  year = {2017},
  doi = {10.1016/j.socec.2017.03.002}
}

@article{maniscalco2012signal,
  title = {A Signal Detection Theoretic Approach for Estimating Metacognitive Sensitivity from Confidence Ratings},
  author = {Maniscalco, Brian and Lau, Hakwan},
  journal = {Consciousness and Cognition},
  volume = {21},
  number = {1},
  pages = {422--430},
  year = {2012},
  doi = {10.1016/j.concog.2011.09.021}
}

@article{moore2008trouble,
  title = {The Trouble with Overconfidence},
  author = {Moore, Don A. and Healy, Paul J.},
  journal = {Psychological Review},
  volume = {115},
  number = {2},
  pages = {502--517},
  year = {2008},
  doi = {10.1037/0033-295X.115.2.502}
}

@article{mccoy2024embers,
  title = {Embers of autoregression show how large language models are shaped by the problem they are trained to solve},
  author = {McCoy, R. Thomas and Yao, Shunyu and Friedman, Dan and Hardy, Mathew D. and Griffiths, Thomas L.},
  journal = {Proceedings of the National Academy of Sciences},
  volume = {121},
  number = {41},
  pages = {e2322420121},
  year = {2024},
  doi = {10.1073/pnas.2322420121}
}

@inproceedings{zhou2024reliable,
  title = {Relying on the Unreliable: The Impact of Language Models' Reluctance to Express Uncertainty},
  author = {Zhou, Kaitlyn and Hwang, Jena D. and Ren, Xiang and Sap, Maarten},
  booktitle = {Proceedings of the 62nd Annual Meeting of the Association for Computational Linguistics (Volume 1: Long Papers)},
  pages = {3623--3643},
  year = {2024},
  publisher = {Association for Computational Linguistics},
  doi = {10.18653/v1/2024.acl-long.198}
}

@article{li2026metacognitive,
  title = {Modeling the joint impact of human and {AI} metacognitive sensitivity on human--{AI} collaboration},
  author = {Li, ZhaoBin and Steyvers, Mark},
  journal = {Journal of Mathematical Psychology},
  volume = {129},
  pages = {102988},
  year = {2026},
  doi = {10.1016/j.jmp.2026.102988}
}

@article{lee2004trust,
  title = {Trust in automation: Designing for appropriate reliance},
  author = {Lee, John D. and See, Katrina A.},
  journal = {Human Factors},
  volume = {46},
  number = {1},
  pages = {50--80},
  year = {2004},
  doi = {10.1518/hfes.46.1.50_30392}
}

@inproceedings{kelly2023capturing,
  title = {Capturing Humans' Mental Models of {AI}: An Item Response Theory Approach},
  author = {Kelly, Markelle and Kumar, Aakriti and Smyth, Padhraic and Steyvers, Mark},
  booktitle = {Proceedings of the 2023 ACM Conference on Fairness, Accountability, and Transparency (FAccT '23)},
  pages = {1723--1734},
  year = {2023},
  publisher = {ACM},
  doi = {10.1145/3593013.3594111}
}

@article{depaoli2017wambs,
  author = {Depaoli, Sarah and van de Schoot, Rens},
  title = {Improving Transparency and Replication in Bayesian Statistics: The {WAMBS}-Checklist},
  journal = {Psychological Methods},
  year = {2017},
  volume = {22},
  number = {2},
  pages = {240--261},
  doi = {10.1037/met0000065}
}

@article{lebuda2024no,
  title = {No strong support for a Dunning--Kruger effect in creativity: analyses of self-assessment in absolute and relative terms},
  author = {Lebuda, Izabela and Hofer, Gabriela and Rominger, Christian and Benedek, Mathias},
  journal = {Scientific reports},
  volume = {14},
  number = {1},
  pages = {11883},
  year = {2024},
  publisher = {Nature Publishing Group UK London},
  doi = {https://doi.org/10.1038/s41598-024-61042-1}
}

@article{fernandes2026explaining,
  title = {Explaining Too Much? Understanding How Large Language Model Reasoning Traces Influence Performance and Metacognition},
  author = {Fernandes, Daniela and Buschek, Daniel and Tankelevitch, Lev and Kosch, Thomas and Welsch, Robin},
  journal = {arXiv preprint arXiv:2605.25856},
  year = {2026}
}
\clearpage
\appendix

\newpage

\appendix
\section*{Appendices}
The appendices document item generation (Appendix~\ref{app:generators}), study procedure and questionnaires (Appendix~\ref{app:study_procedure}), AI model benchmarking (Appendix~\ref{app:harness}), analysis specifications (Appendix~\ref{app:analysis_details}), supplementary results and robustness (Appendix~\ref{app:supplementary_results}), and computational diagnostics including WAMBS (Appendix~\ref{app:bayesian_diagnostics}).
\label{sec:appendix}

\section{Item generation}
\label{app:generators}
The following subsections document the construction rules, difficulty parameters, and validity constraints for each task generator.

\subsection{Item Difficulty}
\label{app:item_difficulty}

We selected designed difficulty levels to vary performance within and across blocks (Table~\ref{tab:battery}). Difficulty labels are generator-specific, not empirically equated across tasks.

\begin{table}[h!]
\centering
\caption{Composition of the administered 40-item battery. Difficulty labels are
the generator's designed levels, not empirical difficulties.}
\label{tab:battery}
\small
\begin{tabularx}{\linewidth}{@{}l r l X@{}}
\toprule
Task type & $n$ & Difficulty mix & Response format \\
\midrule
Matrix reasoning        & 10 & 2 medium, 4 hard, 4 expert &  Multiple choice (8 options) \\
Mental rotation         & 10 & 2 medium, 4 hard, 4 expert & Multiple choice (2 options, same/different object) \\
Syllogistic reasoning   & 10 & 1 easy, 4 medium, 5 hard\textsuperscript{a} & Multiple choice (3 options, valid/invalid/can not be determined) \\
Letter-string analogies & 10 & 3 hard, 4 expert, 3 combination & Free-text symbol sequence \\
\bottomrule
\end{tabularx}
\vspace{2pt}
\footnotesize\textsuperscript{a}Presented in six scenarios, with two items carrying three questions each and four carrying one.
\end{table}

\subsection{Matrix reasoning}
\label{app:gen:mat}

Each attribute that varies in an item is governed by one of four relation types after~\citet{carpenter_what_1990} and \citet{Matzen2010}:

\begin{itemize}
  \item constant: the same value in all three cells of a row, differing
        between rows
  \item progressive: the value cycles across the columns, in the same
        order in every row
  \item unique: the value follows a diagonal cycle, so that the cell in
        row $r$ and column $c$ takes the value at index $(r+c) \bmod 3$
  \item distribute-three: each of the three values appears exactly once
        per row, in an order drawn at random for each row
\end{itemize}

Difficulty is set by how many of the attributes vary at once and by how many of them use distribute-three, as shown in Table~\ref{tab:matladder}.

\begin{table}[h!]
\centering
\caption{Difficulty levels for matrix items.}
\label{tab:matladder}
\small
\begin{tabular}{l c l l}
\toprule
Level & Attributes varying & Relations drawn from & Additional constraint \\
\midrule
easy   & 2       & classic only & none \\
medium & 3       & classic only & none \\
hard   & 4       & all four     & at least one distribute-three \\
expert & 4\textsuperscript{a} & all four & at least two distribute-three \\
\bottomrule
\end{tabular}

\vspace{2pt}
\footnotesize\textsuperscript{a}At the expert level all five attributes are sampled, but because
orientation is always among them, the shape rule is removed by the orientation--shape exclusion
below and no unused attribute remains to replace it, so every expert item varies the four
non-shape attributes (count, fill, size, orientation), with shape fixed.
\end{table}

Two rendering constraints prevent visually ambiguous options. Orientation rules are restricted to shapes on which orientation is visible (triangles, with the three orientation values separated by at least $25\%$ of the $120^{\circ}$ rotational period); a sampled rule set pairing orientation with a shape rule swaps the shape rule onto an unused attribute, or drops it at the expert level (Table~\ref{tab:matladder}). Distractors are deduplicated on appearance rather than description: a visual key collapses orientation variants that a shape's rotational symmetry renders indistinguishable, and any distractor sharing the correct answer's visual key, or duplicating another distractor's, is removed.

Element size is constant across counts, keeping count and size orthogonal cues, and grid and option images are rendered from shared geometry constants, so a shape is pixel-identical between question and options.

\subsection{Mental rotation}
\label{app:gen:rot}

Figure~A is always shown at its canonical orientation. How Figure~B is produced
depends on the condition. In the \emph{depth} condition the object itself is
rotated in three dimensions. In the \emph{picture-plane} condition the object is
not rotated at all and the camera is rolled by the same angular disparity,
replicating the picture-plane condition of the original paradigm. Both conditions are
crossed with the four difficulty levels in Table~\ref{tab:rotladder}.

\begin{table}[h!]
\centering
\caption{Difficulty levels for mental rotation. The rotation axis applies to the depth condition only, and it switches with the angle, so no axis effect is separable from an effect of angular disparity; with the camera azimuth fixed at $45^{\circ}$ the switch is not itself a difficulty manipulation.}
\label{tab:rotladder}
\small
\begin{tabular}{l c l}
\toprule
Level & Angular disparity & Rotation axis (depth condition) \\
\midrule
easy   & $40^{\circ}$  & world $Y$ \\
medium & $80^{\circ}$  & world $Y$ \\
hard   & $120^{\circ}$ & world $X$ \\
expert & $160^{\circ}$ & world $X$ \\
\bottomrule
\end{tabular}
\end{table}

Every figure is ten cubes in four segments, giving the three right-angled elbows of the original form; each new cube must be adjacent to exactly one prior cube, which makes closed rings impossible, and base figures are screened for chirality against all 24 proper lattice rotations, so a ``different'' pair is never an achiral figure some rotation could match. Three implementation choices protect the construct: rotation is applied to polygon face vertices as exact floats rather than integer cube centers, so rotated figures carry no lattice-rounding artifacts; the camera's viewing direction never changes (elevation $25^{\circ}$, azimuth $45^{\circ}$), stationary in the depth condition as in the original method and adding only a roll about the fixed viewing axis in the picture-plane condition; and face shading is reassigned after rotation to the nearest of the six canonical shades by world-space normal, so shading follows orientation rather than identity and cannot signal handedness.

\subsection{Syllogistic reasoning}
\label{app:gen:syl}

Each item is a two-premise categorical argument with a stated conclusion, judged as \emph{Valid} (necessarily true given the premises), \emph{Invalid} (necessarily false), or \emph{Cannot be determined} (neither entailed nor refuted, or the premises are inconsistent). The generator dresses each syllogism as a record-keeping narrative in one of six mundane domains (a bakery, a school, a sports club, a library, an animal shelter, a garden center), rewriting the premises so that the canonical quantifier words (\emph{all}, \emph{no}, \emph{some}) are replaced by prose paraphrases from which the quantifier must be recovered, and weaving in neutral distractor sentences.

Difficulty tiers use the Johnson--Laird mental-model count \cite{JohnsonLaird1984}: the number of distinct subject--predicate relationships permitted by the premises, computed by an independent solver. Easy, medium, and hard items permit one, two, and three models, respectively. Historical hand-estimated human accuracies were not used as validation data.

Three-question scenarios use nine category phrases in three term-disjoint triples, with no category shared between premise pairs. Disjointness is asserted at generation time. Shared narrative and interaction context can still produce correlated response errors.

Difficulty is only a tiebreaker among equally label-useful candidates. All 18 hard single-question candidates have the answer \emph{Cannot be determined}, and only two of 48 forms yield a provably false conclusion. The selection procedure therefore enforces at least one valid and one invalid item to prevent a constant response from solving the battery.

\subsection{Letter-string analogies}
\label{app:gen:ls}
Every item presents the alphabet ordering, two worked examples of one transformation, and one target sequence to complete.
The two alphabets are a permutation of the $26$-letter Latin alphabet in which $20$ of the $26$ positions are deranged (six letters keep their alphabetic position) and a $15$-symbol non-alphabetic set (\textit{> * + < ! @ \$ ) \& = : - ( \% \textasciitilde}). Items are built from four single transformations --- add-letter, fix-alphabet, sort, and a simultaneous successor-and-predecessor operation on opposite ends --- and three ordered pairs of transformations: remove-redundant + add-letter, remove-redundant + sort, and sort + add-letter.

Difficulty is built from the generalization axes of~\citet{Lewis2024}: a step size greater than one, a longer sequence, and grouping, meaning each symbol is displayed twice so the sequence must be parsed before the rule can apply. Each item records how many axes are active at once, counted relative to a baseline of step one, length five and no grouping --- the \emph{medium} level in Table~\ref{tab:lsladder}, which was not administered. 
The alphabet is not one of the counted axes. It changes between levels alongside them --- hard moves to the symbol set as the step rises to two, and expert returns to the $26$-letter alphabet as length and grouping are added --- so no effect of the alphabet can be separated from the axes it moves with.
Table~\ref{tab:lsladder} gives the parameters of each level.

Responses are typed as free text and scored server-side by whitespace- and case-normalized exact match against the key: runs of whitespace collapse to single spaces and case is ignored, so spacing and capitalization cannot make a correct sequence wrong, and the two documented alternate answers (see below) are also accepted.

\begin{table}[tb]
\centering
\caption{Parameters of the letter-string levels. The \emph{medium} level is
shown as the baseline against which generalization axes are counted, but was not
administered.}
\label{tab:lsladder}
\small
\begin{tabular}{l l c c c c}
\toprule
Level & Alphabet & Step & Base length\textsuperscript{a} & Grouping & Axes active \\
\midrule
medium (baseline) & $26$-letter & 1 & 5 & no  & 0 \\
hard              & $15$-symbol & 2 & 5 & no  & 1 \\
expert            & $26$-letter\textsuperscript{b} & 2 & 9 & yes & 3 \\
combination       & mixed\textsuperscript{c} & 2 & 5 & yes & 3\textsuperscript{c} \\
\bottomrule
\end{tabular}

\vspace{2pt}
\footnotesize\textsuperscript{a}An \emph{element} is one position in the sequence, that is, one symbol of the alphabet. Base length is the length of the sequence the generator draws. The strings actually shown are derived from it, so how many elements they contain depends on the transformation: an add-letter item shows one element fewer before the arrow than after, a remove-redundant item one more. Shown strings run from $4$ to $6$ elements at the medium, hard and combination levels and from $7$ to $10$ at expert. Where grouping is used every element is printed twice, so \texttt{[k k r r e e g g]} is four elements and eight printed symbols.

\footnotesize\textsuperscript{b}The $15$-symbol set cannot support the expert level: a nine-element step-two sequence spans $17$ alphabet positions and the symbol set has only $15$. At the shorter levels either alphabet fits, so the alphabet varies alongside the axes rather than independently of them, and it is not one of the counted axes.

\footnotesize\textsuperscript{c}The alphabet is drawn per item: two of the three combination items use the $15$-symbol set and one the $26$-letter set. Their third counted generalization axis is the two-rule composition itself rather than one of the axes above.
\end{table}

\paragraph{Ambiguity checks and combination items}
The verifier enumerates a hypothesis set $\mathcal H$: structural rule families with step sizes 1--4, positional swaps and removals, per-position \emph{fix the odd one out} readings, alphabet-order sorting, duplicate removal, arithmetic-progression repair, and literal per-position index shifts. An item is accepted only if every rule consistent with both worked examples yields the intended target answer. A self-check independently re-parses the stored display strings and repeats verification.

Combination items use remove-redundant $+$ sort, remove-redundant $+$ add-letter, or sort $+$ add-letter. For these items, $\mathcal H$ also contains ordered two-operation compositions of the primitive rules. This guarantee is relative to the enumerated hypotheses: content-addressed rules, such as swapping whichever adjacent pair is out of order, may remain undetected.

\paragraph{Accepted alternate answers}
Two combination-level items were genuinely under-determined in a way the verifier did not catch: each worked example needed only one adjacent-pair swap to reach the intended order, so ``sort into alphabet order'' and ``swap the single out-of-order pair'' fit both examples equally well, and only the target distinguishes them. For these two items the answer implied by the narrower rule is also accepted as correct.

\FloatBarrier
\section{Study procedure, interface, and questionnaires}
\label{app:study_procedure}

\subsection{Additional procedure and interface details}
\label{app:procedure_detail}
The consent flow covered the categories of data collected, storage linked only to the Prolific identifier rather than names or contact details, browser-based proctoring, the retention period, and withdrawal rights. Participants unwilling to be proctored were instructed to return the study. Those who consented entered their Prolific identifier rather than their name when the proctoring service requested one, then re-entered the study inside the proctoring frame. The Human+AI tutorial explained how to copy an item into the chat and send stimulus images, gave example prompts, and stated the consultation strategy.

The application used a React front end, Python/Flask back end and PostgreSQL database. Two logged buttons transferred items into the chat: one copied the item and answer options for pasting, excluding the participant's confidence rating; the other sent the stimulus image for matrix reasoning and mental rotation. Response controls stayed active throughout. The one-message requirement was checked at submission, so participants could decide on an answer before consulting the AI model. GPT-5.6 Luna was selected from seventeen candidates screened for response speed, cost and accuracy, and used at low reasoning effort throughout. The application logged item response times, session duration and, in Human+AI, messages sent per item.

\subsection{Questionnaire sections and attention checks}
\label{app:questionnaire}

Four post-task questionnaire sections were administered. \emph{About you} (all participants) covered age, gender, education, profession, frequency of AI-tool use ($5$-point, \emph{Never}--\emph{Every day}), and preferred AI provider and model. \emph{AI consultation strategy} (\textit{AI} only) asked for the percentage of items with interaction beyond the mandated minimum, a four-option description of the typical approach, and behavior when the assistant disagreed with the participant's initial answer --- a self-report companion to the behavioral reliance measure. \emph{AI interaction} (\textit{AI} only) rated perceived helpfulness, trust, and frustration on fully labeled $5$-point scales. \emph{Task feedback} (all) rated the task set's suitability and collected optional strategy descriptions and free-text comments. Full item wording is included with the study materials,  (see the availability note in Section~\ref{sec:method:availability}).

Three attention checks required the instructions to have been read rather than skimmed: one on how the $\pounds100$ bonus is earned, whose strongest distractor is the \emph{other} bonus in the same incentive card; one condition-specific check probing the engagement disclaimer (Human+AI) or the stated size of the top-10 bonus (Human alone); and a standard instructed-response item after the battery (``select the leftmost option''), carried by a condition-appropriate scale. The stated exclusion rule was failure on two or more checks (Section~\ref{sec:method:participants}). Among the 366 analyzed participants, 315 passed all three checks and 51 failed exactly one; none failed two or more. Human alone contributed 162 passing all three and 25 failing one; Human+AI contributed 153 and 26, respectively. Failures concerned the bonus-rule check (40), the instructed-response check (5), and the condition-specific check (6).
The 4 August 2026 before/after attention-filter snapshots separately document nine removed Human alone participants, all with two failed checks and none in the final sample. This is a dated cleaning step, not a complete recruitment-flow count.

\FloatBarrier
\section{AI model benchmarking harness}
\label{app:harness}
 This appendix documents the harness behind the AI-alone reference reported in Section~\ref{sec:method:benchmark}, which contains 4,000 scored item records from 100 complete GPT-5.6 Luna runs at low reasoning effort.

 \paragraph{Prompt format and scoring.}
 The chat-parity runs reproduce participant-facing message text and images without a system prompt, format instruction, or answer prefill. Free-form replies are scored using extracted answers, with Gemini 3.5 Flash Lite and a deterministic bracket-reference fallback. We use the rescored correctness field, not the legacy single-token regular-expression scorer. Twelve Luna letter-string records are flagged for review and remain as recorded in the primary analysis. Correcting the one confirmed extraction error changes the benchmark mean from $27.71$ to $27.72/40$; scoring all twelve correct gives $27.83/40$. Failed calls are scored incorrect. No Luna record is flagged as a refusal. Multi-question syllogism pages share a reply, so item records must not be counted as separate API calls.

 \paragraph{Run bookkeeping.}
 Run identifiers index 100 shuffled presentations of the same 40 items. Every selected Luna run contains all 40 items, with no duplicate run--item keys. Runs are repeated realizations on this fixed battery, not independent samples of items. The benchmark omits instruction/practice context and resets history at each task block. 

\FloatBarrier
\section{Additional analysis specifications}
\label{app:analysis_details}
For paired contrasts, $d_z$ is the mean individual difference divided by its standard deviation. Component-belief regressions use participant-clustered CR1 covariance and $t$ inference with $G-1$ degrees of freedom ($G$ participants). An endpoint sensitivity excludes rows where any report is 10. These frequentist regressions have no priors. Percentile midpoint tests are Holm-adjusted across the two measures separately within each group, and the two between-group tests form a separate family. Supplemental pooled item regressions include AI model performance by group interactions and item-specific random slopes for group.

\paragraph{Bayesian equivalence specifications}
Exploratory equivalence checks used \texttt{brms} Gaussian mean models \cite{burkner2017brms}. Two-group responses were centered on the grand mean and divided by the observed pooled within-group SD; the group coefficient had prior $\mathcal N(0,1)$, the centered intercept $\mathcal N(0,2)$, and each group's log residual SD $\mathcal N(0,1)$. One-sample responses were centered on 50 and divided by their observed SD; their intercept had prior $\mathcal N(0,1)$ and residual SD a half-Student-$t(3,0,1)$ prior. Normal second arguments are SDs. Each posterior mean contrast was standardized by the posterior pooled residual SD (one-sample: residual SD). We report $P(|d|<.30\mid\mathrm{data})$, using $.95$ as the criterion for practical equivalence. The margin was selected post hoc; $.10$ and $.20$ margins and a wider $\mathcal N(0,2)$ effect prior were sensitivity checks. All primary fits used four chains of 20,000 iterations, including 4,000 warmup per chain. These are approximate mean comparisons, not models of endpoint mass or Bayes factors for exact equality.

\FloatBarrier
\section{Supplementary results and robustness}
\label{app:supplementary_results}
The following sections provide engagement summaries, correlations, and robustness analyses supporting the main Results. Refer to the main text for whole-battery and block descriptive tables.

\subsection{Engagement and correlations}
Correlations involving estimation error share an achieved-score term and are not independent evidence for DKE (\autoref{tab:cor_robin}).

\begin{table}[!htp]
\centering
\caption{Correlations among metacognitive measures. $\Delta EP$ = estimate $-$ performance; $\Delta$conf = confidence gap between correct and incorrect items; $\mu$conf = mean item confidence. $^{*}p<.05$, $^{**}p<.01$, $^{***}p<.001$.}
\label{tab:cor_robin}
\footnotesize
\setlength{\tabcolsep}{4pt}
\begin{tabular}{l rrrrr}
\toprule
& $\Delta EP$ & Est. & Perf. & $\Delta$conf & $\mu$conf \\
\midrule
\multicolumn{6}{l}{\textit{Human+AI}} \\
Estimate & 0.72*** &  &  &  &  \\
Performance & -0.64*** & 0.07 &  &  &  \\
$\Delta$conf & -0.15 & -0.27*** & -0.09 &  &  \\
$\mu$conf & 0.31*** & 0.52*** & 0.13 & -0.15* &  \\
AUROC & -0.17* & -0.21** & 0.02 & 0.84*** & -0.07 \\
\midrule
\multicolumn{6}{l}{\textit{Human alone}} \\
Estimate & 0.63*** &  &  &  &  \\
Performance & -0.45*** & 0.40*** &  &  &  \\
$\Delta$conf & -0.27*** & -0.28*** & 0.01 &  &  \\
$\mu$conf & 0.37*** & 0.67*** & 0.34*** & -0.27*** &  \\
AUROC & -0.19** & -0.20** & -0.00 & 0.87*** & -0.23** \\
\bottomrule
\end{tabular}
\end{table}

\begin{table}[!htp]
\centering
\caption{Maximum number of prompts sent to the assistant by a participant, across all items.}
\label{tab:prompts_robin}
\begin{tabular}{lr}
\toprule
Maximum prompts & Participants \\
\midrule
1 & 8 (4\%) \\
2 & 30 (17\%) \\
3 & 44 (25\%) \\
4 & 36 (20\%) \\
5 & 28 (16\%) \\
>5 & 33 (18\%) \\
\bottomrule
\end{tabular}
\end{table}

\subsection{Robustness procedures and sensitivities}
\label{app:robustness_details}
The first part of RQ3 investigated whether the DKE survives controls for regression to the mean and measurement error, that is, whether the pattern remains once participants are no longer ranked on the same score that is subtracted from their estimate \cite{krueger2002unskilled}. Overestimation is an estimate minus a score, and the quartile difference is overestimation in the lowest quartile minus overestimation in the highest. 

We applied four controls. The split-score control adapts the odd/even different-test analysis of \citet{krueger2002unskilled}, who used percentile errors; here the outcome is an absolute-score error against a doubled half-score. The grouping-only variant still shares items with the outcome and is not a fully disjoint control.
The naive analysis ranks and evaluates on all 40 items. The grouping-only variant ranks on the 20 odd items but evaluates on all 40, giving Q1--Q4 contrasts of $11.98$ (Human+AI) and $8.60$ (Human alone). The strict control ranks on odd items and evaluates the estimate minus twice the even score; its results are in Section~\ref{sec:robust_robin}. Figure~\ref{fig:dkeblocks} instead ranks on the other 30 items and evaluates estimation error on the target block's ten items.

Second, following the simulation approach of \cite{nuhfer2016random} and \cite{gignac2020dunning}, who generated data in which self-assessment carried no information about ability beyond noise, we simulated a population in which no participant's latent report bias depended on their ability. Our generator differs as it uses a calibrated logistic-normal ability, a binomial score, and an additive report that is rounded and censored at the scale bounds.
We simulated a population in which no participant's latent report bias depended on their ability (constant additive bias and noise before rounding and censoring), and applied the same quartile analysis to it. Across 1,500 simulations, we find that this specified null produced a quartile difference of $+3.91$ (95\% simulation reference interval $[+2.13, +5.61]$) in the Human+AI group and $+3.55$ ($[+1.32, +5.93]$) in the Human alone group, as compared to observed differences of $+13.58$ and $+8.87$. No simulation in either group reached the observed value, both $p < .001$. The simulated difference is about $29\%$ of the Human+AI difference and $40\%$ of the Human alone difference, so this specified null does not reproduce the observed contrasts.

Third, the split-half IV analysis uses one item half as an instrument for performance on the other, correcting measurement error under the instrument assumptions. The corrected slopes and reliability estimates are reported in Section~\ref{sec:robust_robin}. This adapts \citet{feld2017estimating}, whose empirical instrument was prior grade-point average for exam performance rather than another half of a reasoning battery.

Fourth, we examined curvature and residual spread using the diagnostics of \citet{gignac2020dunning}. These assess a linear, constant-spread account, not a general distinction between statistical and metacognitive explanations. The relation could not be distinguished from linear in either group (Human+AI $p = .902$, Human alone $p = .129$). The Glejser test regresses the absolute residuals of that relation on the score, so that a negative coefficient indicates that estimates scatter less around the relation as the score rises; a flattened relation with constant noise would leave that scatter the same at every score. We find that estimates varied less among high performers in both groups (Human+AI: $-0.117$, $p = .036$; Human alone: $-0.155$, $p = .001$), describing a gradient in between-participant residual spread. We report an unstandardized residual-spread slope rather than the Glejser correlation reported in that paper.

To check the bounded response scale, we fitted a rounded, censored normal response distribution conditional on total score, with a linear latent mean and constant latent SD. \citet{magnus2022statistical} show that a statistical model that takes the boundaries of the scale into account can reproduce the DKE pattern on its own. Our check applies that idea to the residual-spread statistic and is not their specification. Repeating the identical two-stage Glejser statistic on 1,500 simulated datasets gave lower-tail $p=.009$ in Human+AI and $p<.001$ in Human alone. This plug-in test evaluates that specific constant-noise account; uncertainty in its fitted parameters and the observed ability scores is not integrated out.

We used $4{,}000$ participant bootstrap resamples within each group, re-forming quartiles in every resample. For the two-stage residual-spread analyses, both stages were refitted in each resample while holding the odd/even item split fixed \cite{lebuda2024no}

The fixed-split residual-spread slopes were $-0.117$ (95\% CI $[-0.227,-0.008]$) in Human+AI and $-0.116$ ($[-0.201,-0.023]$) in Human alone. In an exploratory sensitivity analysis, adjusting the second stage for session duration and log median response time gave slopes of $-0.164$ ($[-0.280,-0.044]$) and $-0.130$ ($[-0.237,-0.018]$), respectively. These adjustments assess associations conditional on measured timing.

The Human+AI minus Human alone difference in the strict controlled Q1--Q4 contrast was $2.38$ points (95\% participant-bootstrap CI $[-2.14,7.22]$). To avoid distributing potentially dependent questions across halves, we repeated the analysis with questions sharing a syllogism scenario assigned to the same half.
This scenario-preserving split gave Q1--Q4 contrasts of $8.80$ points in Human+AI and $5.96$ in Human alone, with a group difference of $2.84$ (95\% CI $[-2.16,6.82]$). The controlled contrasts therefore remained positive in both groups. Measurement-error-corrected tracking slopes using the scenario-preserving split were $0.104$ (95\% CI $[-0.132,0.345]$) in Human+AI and $0.544$ ($[0.356,0.732]$) in Human alone.

Separately, $60$ random item splits assessed the sensitivity of the residual-spread results to item allocation. These splits balanced task blocks but did not preserve shared syllogism scenarios. Their ranges are exploratory sensitivity summaries, not participant-sampling confidence intervals, and are distinct from the $1{,}500$-draw null-simulation reference intervals.
\

\subsection{Pooled item-level interaction sensitivities}
\label{app:item_sensitivity}
Supplemental pooled specifications with item-specific group contrasts estimated an accuracy interaction of $0.957$ logits per SD, Wald CI $[0.730,1.184]$, $z=8.26$, $p<.001$, and a confidence interaction of $-0.0007$, CI $[-0.0131,0.0117]$, $z=-0.12$, $p=.908$. The confidence specification had a singular random-effects fit and is treated as a sensitivity analysis. These separate interaction tests do not establish a difference between accuracy and confidence slopes. Uncertainty is conditional on the 100-run benchmark.

\FloatBarrier
\section{Computational diagnostics and WAMBS}
\label{app:bayesian_diagnostics}

\subsection{Computational fit checks and priors}
\label{app:computational_checks}

\begin{table}[!ht]
\centering
\caption{Retrospective prior rationale. $\mathcal{N}^{+}$ and $t^{+}$ denote positive-truncated distributions. Computational parameters are on the latent probit-type scale unless stated otherwise; brms outcomes use the observed SD for numerical scaling. Full sensitivity settings and induced response-scale checks are retained in the analysis code.}
\label{tab:wambs_priors}
\footnotesize
\begin{tabularx}{\linewidth}{p{.19\linewidth}p{.28\linewidth}X}
\toprule
Parameter & Primary prior & Rationale / interpretation \\
\midrule
Skill $\theta_i$ & $\mathcal{N}(0,2)$ & Population location reference and continuity with the parent account; inferred jointly, not exogenous ability. \\
Report shifts $b_k,c_k$ & $\mathcal{N}(0,2)$ & Symmetric latent shifts; neither is directly a score-point bias. $c_k$ appears only in the extended specification. \\
Report scale $\sigma_k$ & $\log\sigma_k\sim\mathcal{N}(0,2)$ & Positive divisor allowing compression or expansion; broad, with potentially extreme induced curves. \\
Raw difficulty $d_{kj}$ & $\mathcal{N}(0,1)$, centered by group & Relative block locations sum to zero; induced marginal SD is $\sqrt{3/4}$. \\
Person offset & $u_i=\tau_k z_i$; $z_i\sim\mathcal{N}(0,1)$; $\tau_k\sim\mathcal{N}^{+}(0,1)$ & Partial pooling across repeated reports; repetition does not guarantee identification. \\
brms mean effect & $\mathcal{N}(0,1)$; wider check $\mathcal{N}(0,2)$ & Proper regularization after outcome scaling. Reported $d$ additionally uses the posterior residual SD. \\
brms intercept / SD & Two groups: intercept $\mathcal{N}(0,2)$, each log-SD $\mathcal{N}(0,1)$. One sample: SD $t^{+}_3(0,1)$. & Allows unequal group variances. One-sample mean priors apply to deviations from the percentile midpoint. \\
\bottomrule
\end{tabularx}
\end{table}

For the extended specification, the conditional expected Q1--Q4 overestimation contrast was 12.45 [11.75, 13.12] points in Human+AI and 8.18 [7.37, 9.04] in Human alone. For the total-only specification, the conditional expected Q1--Q4 overestimation contrast was 8.24 [7.24, 9.74] points in Human+AI and 4.85 [3.93, 5.79] in Human alone. The observed contrasts using those same quartiles were 13.58 and 8.87 points, respectively.

These are conditional, in-sample checks, not held-out model comparisons. Figure-linked summaries use 2,000 posterior draws; WAMBS contrast summaries below use all retained draws. The replicated-report ceiling misfit remains reported in the main text.

\subsection{Retrospective WAMBS review}
\label{app:wambs_review}
The completed WAMBS review \cite{depaoli2017wambs} found no substantive change in the reviewed score-level predictions or equivalence conclusions under longer sampling and the tested priors. Correlated latent parameters shifted together, while some predictive ceiling misfit remained.

\paragraph{Scope and sampling.}
We reviewed eight computational fits (extended and total-only, each primary, doubled, tighter-prior, and broader-prior) and 42 equivalence fits (seven outcomes, each with two effect priors, primary/doubled sampling, and two nuisance-prior checks). All used four chains. Primary runs retained $16{,}000$ draws per chain after $4{,}000$ warmup; doubled runs retained $32{,}000$ after $8{,}000$. Numerical diagnostics cover all parameters; visual checks cover population parameters, selected person effects, and standardized contrasts.

\begin{table}[!ht]
\centering
\caption{Adapted WAMBS checklist \cite{depaoli2017wambs}. Reviewed means that evidence was examined, not that every criterion passed. C denotes both computational specifications; E denotes the equivalence fits. Full diagnostic files and the evidence map are retained with the analysis.}
\label{tab:wambs_checklist}
\footnotesize
\begin{tabularx}{\linewidth}{p{.18\linewidth}XX}
\toprule
Check & Computational fits (C) & Equivalence fits (E) \\
\midrule
1. Prior meaning & Reviewed: latent-scale rationale and induced score/report distributions; broad endpoint implications retained. & Reviewed: empirical scaling, proper effect/intercept/SD priors, and induced standardized effects. \\
2. Chain behavior & Reviewed: overlapping traces and rank distributions; numerical diagnostics show no sampler flags. & Reviewed: all seven outcomes and six settings each showed overlapping chains. \\
3. Longer runs & Reviewed: independent runs doubled retained draws; contrast and curve changes were small. & Reviewed: doubled both effect-prior fits per outcome; no equivalence conclusion changed. \\
4. Shape and precision & Reviewed with caveat: unimodal inspected histograms; MCSE and interval-endpoint errors exported. & Reviewed with caveat: unimodal effect histograms; some three-decimal precision targets unmet. \\
5. Autocorrelation & Reviewed: decaying ACFs; total-only Human+AI scales mix more slowly, with adequate ESS. & Reviewed: low short-lag dependence; no thinning used for inference. \\
6. Plausibility and fit & Parameter trade-offs preserve score predictions; ceiling and other predictive misfit remain. & Mean comparisons only; Gaussian fits do not model bounded endpoint mass. \\
7. Covariance priors & Not applicable: no estimated multivariate random-effect covariance matrix. Centered difficulties are not such a matrix. & Not applicable: scalar residual SDs, with no covariance matrix. \\
8. Prior influence & Reviewed: prior/posterior overlays show shifts among correlated parameters; score-level conclusions are stable. & Reviewed: induced priors and effect/residual-prior comparisons; not a claim of prior neutrality. \\
9. Sensitivity & Reviewed: conditional quartile findings stable across the defined prior sets; latent scales are not. & Reviewed: effect and nuisance priors; $.10/.20/.30$ margins reported separately from prior sensitivity. \\
10. Reporting & Completed with limitations: settings, estimands, priors, evidence and unresolved misfit retained. & Completed with limitations: post-hoc margins, CrIs and inconclusive equivalence distinguished. \\
\bottomrule
\end{tabularx}
\end{table}

Across 50 fits, maximum $\hat R=1.00089$, minimum bulk/tail ESS were $8{,}221/13{,}006$, minimum E-BFMI was $.767$, and there were no divergences or maximum-treedepth hits. Doubling sampling changed computational contrast medians by at most $.0055$ points and interval endpoints by $.0167$. Equivalence $d$ medians changed by at most $.00125$ and $P(|d|<.30)$ by $.00388$. Some decimal-level precision targets remained unmet (maximum probability MCSE $.00265$; standardized-effect endpoint MCSE $.00148$), without changing a substantive comparison.

\paragraph{Prior specifications and sensitivity.}
Computational tighter-prior SDs were $1.5$ for skill, $1$ for $b/c$, and $.75$ for $\log\sigma$, half-normal $\tau$, and raw difficulty. Broader SDs were $3$, $3$, $2.5$, $2$, and $1.5$, respectively. Equivalence nuisance checks used two-group intercept SDs $1/3$ and log-residual-SD prior SDs $.5/1.5$; one-sample residual priors were half-normal$(0,1)$ and half-$t_3(0,2)$. Prior-predictive distributions were broad, including substantial endpoint mass; these proper reference priors are not claimed to be neutral.

Across computational priors, extended conditional quartile contrasts ranged from $12.39$ to $12.48$ points in Human+AI and $8.17$ to $8.20$ in Human alone. Human+AI medians for $b$, $\sigma$, and $\tau$ ranged over $4.30$--$8.98$, $6.04$--$11.46$, and $2.81$--$5.49$. The primary $b$--$\sigma$ correlation was $.914$ (extended) and $.995$ (total-only). Such parameter trade-offs leave similar score predictions; they do not require withdrawing the descriptive computational finding.

\paragraph{Consequences for conclusions.}
All tested settings supported practical equivalence at the post-hoc $|d|<.30$ margin for group differences in mental rotation, signed error, and absolute error, and for Human alone's two percentile judgments versus 50. This is equivalence of mean contrasts, not accurate individual judgment. Total performance versus the AI model benchmark remained inconclusive ($P=.932$--$.937$), as did within-block AUROC ($P=.881$--$.886$). Smaller margins remain separate sensitivity analyses.

The review supports computational stability, not the adequacy of every report-distribution assumption. Primary predictions still underrepresent some Human+AI ceiling reports; alternative mean/likelihood specifications were not adopted and remain outside this review. The observed performance advantage and weak self-assessment tracking also have evidence outside this specification. Its stable conditional quartile contrasts do not replace the disjoint-score test, whose between-group interval includes zero. The reviewed settings therefore preserve the main conclusions without establishing a larger controlled DKE or a unique latent mechanism.

\end{document}